\documentclass[useAMS,usenatbib]{mn2e}

\def\Zsol{\hbox{Z$_{\odot}$}}
\def\Msol{\hbox{M$_{\odot}$}}

\usepackage{epsfig,natbib,url,graphicx}
\usepackage{lscape,amsmath,amssymb}
\usepackage{booktabs,threeparttable}
\usepackage{color}
\usepackage{soul} 
\usepackage{fancyvrb}
\usepackage{pifont}

\newcommand{\hii}{H~{\sc ii}}

\newcommand{\heii}{He~{\sc ii}}

\newcommand{\eld}{$N_{\rm e}$}

\newcommand{\elt}{$T_{\rm e}$}

\newcommand{\Sp}{S$^+$}
\newcommand{\Spp}{S$^{2+}$}
\newcommand{\cp}{C$^+$}
\newcommand{\cpp}{C$^{2+}$}

\newcommand{\op}{O$^+$}

\newcommand{\opp}{O$^{2+}$}

\newcommand{\np}{N$^+$}

\newcommand{\nepp}{Ne$^{2+}$}

\newcommand{\foiii}{[O~{\sc iii}]}
\newcommand{\sfoiii}{O~{\sc iii}]}

\newcommand{\foii}{[O~{\sc ii}]}
\newcommand{\fsii}{[S~{\sc ii}]}

\newcommand{\fciii}{[C~{\sc iii}]}

\newcommand{\fnii}{[N~{\sc ii}]}

\newcommand{\fneiii}{[Ne~{\sc iii}]}

\newcommand{\oi}{O~{\sc i}}

\newcommand{\cii}{C~{\sc ii}}

\newcommand{\mgii}{Mg~{\sc ii}}

\newcommand{\civ}{C~{\sc iv}}
\newcommand{\sIii}{Si~{\sc ii}}
\newcommand{\sIiv}{Si~{\sc iv}}
\newcommand{\sfciii}{C~{\sc iii}]}

\newcommand{\feii}{Fe~{\sc ii}}
\newcommand{\feiii}{Fe~{\sc iii}}
\newcommand{\alii}{Al~{\sc ii}}
\newcommand{\aliii}{Al~{\sc iii}}

\newcommand{\hp}{H$^+$}

\newcommand{\ha}{H$\alpha$}
\newcommand{\hb}{H$\beta$}
\newcommand{\hg}{H$\gamma$}

\newcommand{\aap}{A\&A}

\newcommand{\apj}{ApJ}
\newcommand{\apjl}{ApJL}
\newcommand{\apjs}{ApJS}

\newcommand{\araa}{ArA\&A}
\newcommand{\mnras}{MNRAS}
\newcommand{\pasp}{PASP}

\title[CASSOWARY\,20: comparison of metallicity indicators]
{Testing metallicity indicators at $z\sim1.4$ with 
the gravitationally lensed galaxy CASSOWARY\,20
\thanks{Based on observations made with European Southern Observatory (ESO) telescopes 
at the Paranal Observatory under programme 085.A-0179(A).}}
\author[James et al.]{Bethan L. James$^{1}$\thanks{E-mail: bjames@ast.cam.ac.uk}, Max Pettini$^{1}$, 
Lise Christensen$^2$, Matthew W.  Auger$^{1}$, 
\newauthor George D. Becker$^{1}$, Lindsay J. King$^3$, Anna M. Quider$^1$,
Alice E. Shapley$^4$,
\newauthor  and Charles C. Steidel$^5$\\
$^{1}$Institute of Astronomy, Madingley Road, Cambridge, CB3 0HA\\
$^{2}$Dark Cosmology Centre, Niels Bohr Institute, Copenhagen University, 
Juliane Maries Vej 30, D-2100 Copenhagen \O, Denmark\\
$^{3}$ Department of Physics EC 36, University of Texas at Dallas, 800 West Campbell Road,
Richardson, Texas 75080-3021, USA\\
$^4$ Department of Physics and Astronomy, University of California, Los Angeles, CA 90095-1547, USA\\
$^5$ California Institute of Technology, Mail Stop 105-24, Pasadena, CA 91125, USA\\
}
\begin{document}

\date{Accepted 2014 February 10. Received 2014 February 10; in original form 2013 September 20}

\pagerange{\pageref{firstpage}--\pageref{lastpage}} \pubyear{2013}

\maketitle

\label{firstpage}

\begin{abstract}
We present X-shooter observations of CASSOWARY\,20 (CSWA\,20),
a star-forming  (SFR\,$ \sim6$\,\Msol~yr$^{-1}$)
galaxy at $z = 1.433$, magnified by a factor of 11.5
by the gravitational lensing produced 
by a massive foreground galaxy at $z = 0.741$. 
We analysed the integrated physical properties
of the \hii\ regions of CSWA\,20 
using temperature- and density-sensitive emission lines.
We find the abundance of oxygen to be $\sim 1/7$ of solar,
while carbon is $\sim 50$ times less abundant than in the Sun.
The unusually low C/O ratio may be an indication of a particularly
rapid timescale of chemical enrichment.
The wide wavelength coverage of X-shooter gives us access
to five different methods for determining the metallicity
of CSWA\,20, three based on emission lines from \hii\
regions and two on absorption features formed in the atmospheres
of massive stars. All five estimates are in agreement, within 
the factor of $\sim 2$ uncertainty of each method.
The interstellar medium of CSWA\,20 only partially
covers the star-forming region as viewed from our direction;
in particular, absorption lines from neutrals and first ions
are exceptionally weak. We find evidence for large-scale
outflows of the interstellar medium (ISM) with speeds of up
750\,km~s$^{-1}$, similar to the values measured
in other high-$z$ galaxies sustaining 
much higher rates of star formation.
\end{abstract}


\section{Introduction}
Constraining the degree of metal enrichment in the Universe 
is of paramount importance for our understanding of galaxy 
formation and evolution.  
The metallicity of a galaxy reflects its evolutionary state,
and the most abundant chemical elements
regulate many of the astrophysical processes 
that drive star formation and galaxy evolution: 
cooling, formation of dust and molecules, the opacity and 
energy transport within stellar atmospheres and the synthesis 
of new nuclei within stars.

In galaxies that are actively forming stars, element abundances 
have mostly been determined by modelling the relative
strengths of emission lines from H\,\textsc{ii} regions, 
including weak, temperature-sensitive transitions such as 
\foiii~$\lambda$4363.44 and \fnii~$\lambda$5756.24
(vacuum wavelengths).
This is often referred to as the `direct', or `$T_{\rm e}$', method.
Unfortunately, these auroral lines become increasingly 
difficult to detect in distant, faint galaxies due to the 
limited sensitivity of current instrumentation compounded
with the strong sky background at near-infrared wavelengths,
where the lines fall at redshifts $z \gtrsim 1.2$.
Furthermore, 
most galaxies at $z>1$ with measurements of rest-frame optical 
emission lines are fairly luminous, and, accordingly, not 
extremely metal poor. Since temperature-sensitive auroral 
features decline in strength strongly with increasing 
metallicity, these features are intrinsically weak in the 
majority of high-redshift galaxies studied to date.


\begin{figure*}
\includegraphics[scale=0.47]{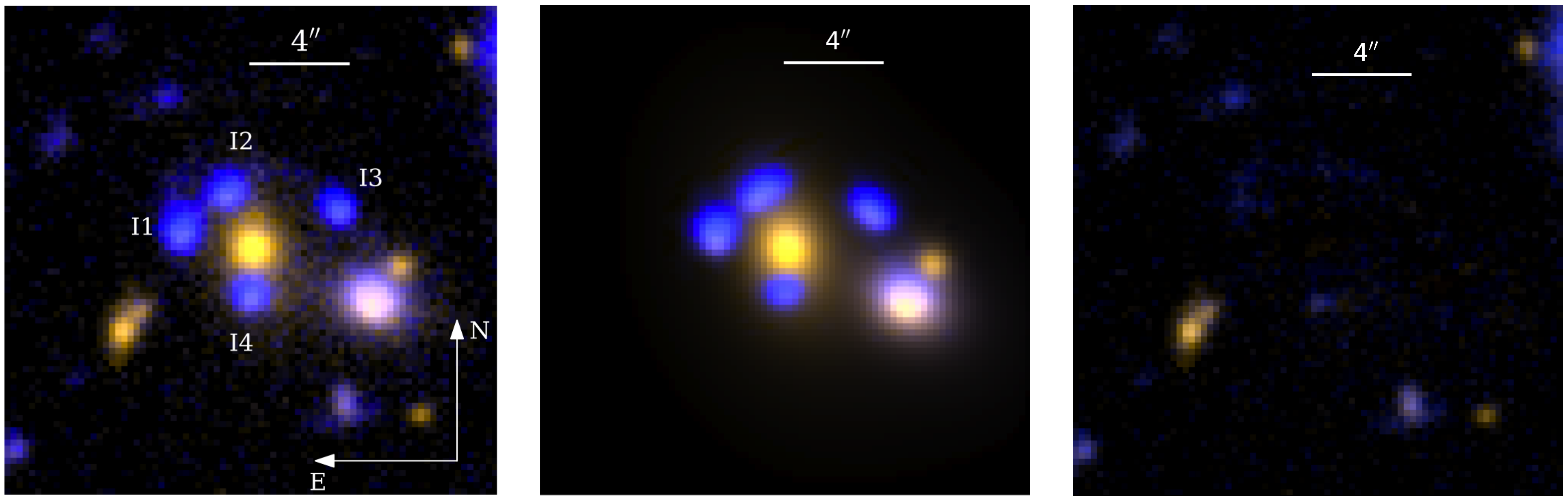}
\caption{\textit{Left:} Colour composite of $g$ and $i$ 
images of the CSWA\,20 lens system. 
The images were obtained with the Auxiliary Port camera
of the William Herschel Telescope (WHT) on La Palma; the exposure times
were 300\,s in each band and the seeing was $\sim 0.8$\,arcsec
FWHM.
Superposed on the image is an outline of the entrance slit
of X-shooter, at the two orientations used (Table~\ref{tab:obs}).~
\textit{Centre:} Image-plane reconstruction of  CSWA\,20 
with the lensing model described in Section~\ref{sec:LensMod}.
Photometry of the system is collected in Table~\ref{tab:mags}.~\textit{Right:} Residuals from subtracting the model (centre) from the data (left).}
\label{fig:CSWA20}
\end{figure*}

In such cases, we are forced to rely on metallicity `indicators' 
which involve the ratios of 
strong emission lines. The most widely used of these  
`strong-line', or `empirical', methods  is the
$R_{23}$ index which relates the abundance of oxygen 
to the  ratio of (\foiii~$\lambda\lambda 5008.24,4960.30$ + 
\foii~$\lambda\lambda 3727.09, 3729.88$)  to \hb$\lambda 4862.69$
\citep{Pagel:1979}. Other combinations of emission lines
have been proposed and extensively used at high redshifts
\citep[e.g.][]{Maiolino:2008};
among these, the $N2$ index (\fnii$\lambda 6585.23$/H$\alpha \lambda 6564.61$)
of \citet{Pettini:2004} has turned out to be particularly useful
for obtaining  a rough measure of a galaxy's metallicity
at redshifts $z \sim 2.3$, where observations from the ground
can capture all the main nebular emission lines in the 
near-infrared atmospheric transmission windows.
This coincidence is especially convenient given that 
$z \sim 2.3$ corresponds to the epoch when  
the Universal star-formation rate was at its peak 
\citep{Madau:2014}.

While these and other strong-line indices have been calibrated 
with reference to abundances measured by the `direct' method,
it is a matter of concern that different calibrations are not
always in agreement and can yield widely different values
of the oxygen abundance \citep{Kewley:2008}.
To make matters worse, it is now clear 
that the physical conditions of the 
most actively star-forming galaxies at $z \gtrsim 2$
are significantly more extreme (higher densities of 
matter and photons, and harder radiation fields) 
than found locally (Hainline et al. 2009; Steidel et al. in preparation).
It is therefore questionable to what extent
the local empirical metallicity indicators
can be reliably applied to high-$z$ galaxies.

In order to make progress on some of these issues,
it would be highly desirable to compare, for the
same galaxy, element abundances
derived with different methods.
At high redshift, there have so far been
only a handful of cases where the `direct' and
`empirical' methods for measuring the oxygen abundance
could be compared
\citep{Villar:2004,Yuan:2009, Erb:2010, Christensen:2012b}.
Generally, the two sets
of values of O/H have been found to agree
within the errors, although in one case
\citep{Yuan:2009} the detection of 
\foiii~$\lambda 4364.44$ in a gravitationally lensed
galaxy at $z = 1.705$ led to a downward revision by a 
factor of about four in the oxygen abundance 
compared to the value estimated from the $R_{23}$ index.


\begin{table}
\begin{center}
\caption{$g$  and $i$  magnitudes of CSWA\,20.$^{\rm a}$}
\begin{tabular}{ccc}
\hline\hline
Image & $g^{\rm b}$ & $i^{\rm c}$ \\
\hline
I1 & 21.91 & 21.87 \\
I2 & 22.25 & 22.32 \\
I3 & 22.49 & 22.49 \\
I4 & 22.47 & 22.42 \\
Total & 20.75 &  20.74  \\
\hline
\end{tabular}
\end{center}
$^{\rm a}$ Measured from the WHT images and lensing model 
shown in Figure~\ref{fig:CSWA20}.\\
$^{\rm b}$ Systematic error on the zero-point of the photometry is $\pm 0.1$\,mag.\\
$^{\rm c}$ Systematic error on the zero-point of the photometry is $\pm 0.17$\,mag.\\
\label{tab:mags}
\end{table}%

The limitations in the use of nebular abundance diagnostics
at high redshifts have motivated the exploration of  
alternative metallicity measures based on spectral
features in the rest-frame ultraviolet (UV) range, including
interstellar absorption lines \citep{Pettini:2002},
P-Cygni emission-absorption features arising
from the winds of the most luminous early-type
stars \citep[][and references therein]{Leitherer:2010},
and photospheric absorption lines from OB stars
[\citet{Rix:2004} and references therein;
see also \citet{Maraston:2009}].
A comparison of these different methods in the very
few cases where it was possible can be found in
\citet[][]{Pettini:2006}; see also \citet{Halliday:2008}
and \citet{Sommariva:2012}.
These UV-based methods are still in the
development stage and are currently  
less accurate than the emission line diagnostics.
Some of the difficulties in their use
are: (i) the fact that some of the photospheric 
features are blends of lines from different ions, 
(ii) the dependence of the stellar wind lines on the 
age of the starburst and on the initial mass function (IMF),
and (iii) complications in separating stellar and interstellar
contributions \citep[]{Crowther:2006b}.
However,
they do offer the hope of determining 
metal abundances of galaxies out to $z\sim7$, 
where the emission lines from \hii\, regions are 
redshifted beyond the reach of ground-based telescopes
(although the UV diagnostics require detecting the stellar
continuum, which is beyond current capabilities 
for galaxies at  $z \gtrsim 5$).

The ability to detect weak emission lines and, more generally,
to compare different metallicity indicators, can be greatly
facilitated by gravitational lensing which makes it possible 
to record the spectra of high-$z$ galaxies at higher signal-to-noise
ratio (S/N) and resolution than would otherwise be the case.
For this reason, most of the examples cited above refer to 
gravitationally lensed galaxies.

While such cases are still relatively rare, their numbers 
have increased substantially in recent years through dedicated searches in
large sky surveys, such as the Sloan Digital Sky Survey (SDSS) ---
eg. \citet[][and references therein]{Kubo:2009};
see also \citet{Richard:2011} and \citet{Stark:2013} for other recent examples.
The CAmbridge Sloan Survey Of Wide ARcs in the skY (CASSOWARY)
targets multiple, blue companions around 
massive ellipticals in the SDSS photometric catalogue 
as likely candidates for wide separation gravitational lens systems
\citep[][]{Belokurov:2009}.
The twentieth source in the CASSOWARY 
catalogue\footnote{The catalogue is available at:\\
http://www.ast.cam.ac.uk/ioa/research/cassowary/},
CSWA\,20, 
was confirmed by \citet{Pettini:2010}
to be four images of the same 
$z_{\rm em}=1.433$ star-forming galaxy
separated by $\sim 3$--5\arcsec\, by the gravitational 
potential of a foreground $z_{\rm abs}=0.741$ massive red galaxy
(see Figure~\ref{fig:CSWA20}).

Following these initial observations, we targeted CSWA\,20
in a dedicated programme using the X-shooter spectrograph 
on the Very Large Telescope (VLT) of the European Southern 
Observatory (ESO), with the aim of recording the galaxy spectrum at
sufficient resolution and S/N for a detailed study of its physical
properties and, in particular, testing the consistency of different
metallicity measures. The results of this work are presented here.

The organisation of the paper is as follows. 
In Section~\ref{sec:Obs}  we describe the X-shooter observations 
and data reduction; we also give a brief description of the 
lensing model.
The full CSWA\,20 X-Shooter spectrum is presented in 
Section~\ref{sec:CSWA_spectrum}, 
along with our analysis of its emission and absorption lines.  
Sections~\ref{sec:abund} and \ref{sec:indicators} 
deal with the measurements of element abundances
by different methods; the results are 
compared in Section~\ref{sec:discuss}.
The main conclusions of this work are summarised in
Section~\ref{sec:summary}.

\section{Observations and Data Reduction}
\label{sec:Obs}
\subsection{Observations}

The X-shooter spectrograph \citep{Vernet:2011}
delivers medium resolution spectra
obtained simultaneously from three separate arms: 
UVB (300--550\,nm), VIS (550--1015\,nm)  and NIR (1025--2400\,nm). 
Our observations of CSWA\,20 used entrance slits of 
1.0\,arcsec (UVB), 0.9\,arcsec (VIS) and 0.9\,arcsec (NIR),
with corresponding resolving powers of $R=\lambda/\Delta\lambda\sim 4400$, 7500 and 5300, 
sampled with $\sim 5.5$, 7.1 and 4.2 pixels respectively 
(after on-chip binning by a factor of two on the UVB and VIS detectors). 

Two orientations of the spectrograph slits on the sky were used,
the first at position angle PA\,$= -53^\circ$ to record
simultaneously the spectrum of images I1 and I2 
(see Figure~\ref{fig:CSWA20}), and the second at 
PA\,$= -46^\circ$ covering I3 and I4. 
The observations used a nodding along the slit approach, 
with an offset of 4\,arcsec between individual exposures;
typically each observation consisted of four 900\,s exposures in 
a ABBA pattern.
A journal of observations is given in Table~\ref{tab:obs}.
Although our plan was to divide the allocated exposure time
equally between the two slit settings, 
this service programme was terminated by ESO 
before it had been completed. Consequently, the
spectra secured are mostly those of images I1 and I2 (see Table~\ref{tab:obs}).
 

\begin{table}
\caption{Journal of observations.}
\begin{center}
\begin{tabular}{lccc}
\hline\hline
Date (UT) & Exposure time (s) & Slit PA($\deg$) & Target \\
\hline
2011 May 07 & 4 $\times$ 900 & 53 & I1 \& I2 \\
2011 July 29 & 4 $\times$ 900 & 53 & I1 \& I2 \\
2011 June 27& 4 $\times$ 900 & 53 & I1 \& I2 \\
2011 May 29& 4 $\times$ 900 & 53 & I1 \& I2 \\
2011 May 23 & 4 $\times$ 900 & 53 & I1 \& I2 \\
2011 May 10  & 4 $\times$ 900 & 53 & I1 \& I2 \\
2011 April 29 & 4 $\times$ 900 & 53 & i1 \& i2 \\
2011 April 09 & 8 $\times$ 900 & 53 & I1 \& I2 \\
2011 May 07 & 2 $\times$ 600 & 46 & I3 \& I4 \\
2011 August 03 & 4 $\times$ 900 & 46 & I3 \& I4 \\
\hline
\label{tab:obs}
\end{tabular}
\end{center}
\end{table}

\subsection{Data Reduction}
\label{sec:DataRed}

The X-shooter spectra were reduced with a suite of customised
routines designed to optimally subtract the sky background
and extract one-dimensional spectra while minimising the noise
\citep{Kelson:2003, Becker:2009}. 
For each observation listed in Table~\ref{tab:obs}, 
we extracted separately the spectrum of each image on the slit;
this resulted in 84 individual 1-D spectra for each of the three
arms of X-shooter. Of these, $\sim 80$ were usable
(in four cases one of the images fell too close to the slit edge 
to allow accurate sky subtraction).

Wavelength calibration used reference spectra of a Th-Ar
hollow-cathode lamp, with small adjustments to the wavelength
solution, when required, to match the wavelengths of 
emission lines from the night sky.
At the S/N ratio of our data, no significant differences
were found between the spectra of the different images,
and indeed 
none are expected from the 
reconstruction of the source (see Section~\ref{sec:LensMod}).
Therefore we combined all 80 spectra with
S/N weighting to produce a  total spectrum in each arm.
From the rms deviations of the data in regions free of obvious
spectral features we measure S/N\,$\simeq 8$, 15 and 5 
for the final spectrum in the UVB, VIS  and NIR arms respectively.

The spectra were not corrected for telluric absorption 
because no emission or absorption lines of interest were 
affected.
For flux calibration, we scaled the X-shooter final VIS arm
spectrum to match the $i$-band magnitude measured from the
WHT images (Figure~\ref{fig:CSWA20} and Table~\ref{tab:mags})
and the $r$-band magnitude measured from 
the X-shooter acquisition images.
Using spectral overlap regions, the UVB and NIR spectra were 
then scaled accordingly to the flux-calibrated VIS spectrum.  
Uncertainties resulting from flux calibration were estimated to be 
$\sim 10$\%.

\subsection{Lensing Model}
\label{sec:LensMod}

The gravitational lens model was determined by fitting a singular
isothermal ellipsoid (SIE) mass model with external shear to the WHT $g$ and
$i$ imaging data shown in Figure~\ref{fig:CSWA20}, 
following the procedure outlined in
Stark et al. (2013) and Auger et al. (2013). The source was modelled as a
single Sersic component; the right panel in Figure 1 shows the
reconstructed image of the system. We find that our model agrees well with the model from Pettini et al. (2010), and in particular we note that our derived SIE velocity dispersion of 496\,km~s$^{-1}$ is in excellent agreement with the spectroscopic velocity dispersion of the lensing galaxy. The mass distribution traces the light well, as each has a flattening of $\sim 0.7$, and we find a significant external shear of $0.18$ oriented at a position angle of 18\,degrees.

We find the total lensing magnification to be $f = 11.5$, $\approx 2-4$
times higher than the initial approximate estimate given by Pettini et al. (2010). This value of $f$ is very precisely determined within the context of our model (the statistical uncertainty from our modelling is less than 1\%). However, although our lensing model reproduces the CSWA\,20 configuration very well 
(see Figure~\ref{fig:CSWA20}), it is not unique. 
Furthermore, as is the case with most lensing models, there may be 
significant systematic uncertainties due to differential magnification of structure 
within the source. Indeed, the magnifications inferred 
from the $g$ and $i$ bands differ 
by $10\%$, possibly signalling differential magnification, 
and we therefore impose a systematic uncertainty of $\pm10\%$ on the 
fluxes used throughout our analysis.


\begin{table}
\begin{center}
\caption{Emission line fit parameters.}
\begin{tabular}{ccc}
\hline\hline
Parameter & Narrow & Broad\\
\hline
\multicolumn{3}{l}{High S/N lines:} \\
\cline{1-1}
$z$ & $1.433485$           &  $1.433074$ \\
       & $\pm 0.000004$ &  $\pm 0.000023$\\
Velocity  offset$^{\rm a}$ (km~s$^{-1}$) & 0.0 & $-51 \pm 3$ \\
$\sigma$(km~s\,$^{-1}$) & $34.9\pm 0.7$ & $109 \pm 3$\\
\hline
\multicolumn{3}{l}{\mgii\,$\lambda\lambda$2796,2803:} \\
\cline{1-1}
$z$                     & 1.43377                   & 1.43335 \\
                           & $\pm 0.00002$     &  $\pm 0.00005$\\
Velocity offset$^{\rm a}$ (km~s$^{-1}$) & $+35 \pm 3$ & $-17 \pm 6$ \\
$\sigma$(km~s\,$^{-1}$) & $40 \pm 2$ & $108 \pm 5 $\\
\hline
\label{tab:params}
\end{tabular}
\end{center}
$^{\rm a}$ Relative to $z = 1.433485$, taken to be the systemic redshift.\\
\end{table}

\section{The Spectrum}
\label{sec:CSWA_spectrum}


\begin{figure*}
\includegraphics[angle=180,scale=0.925]{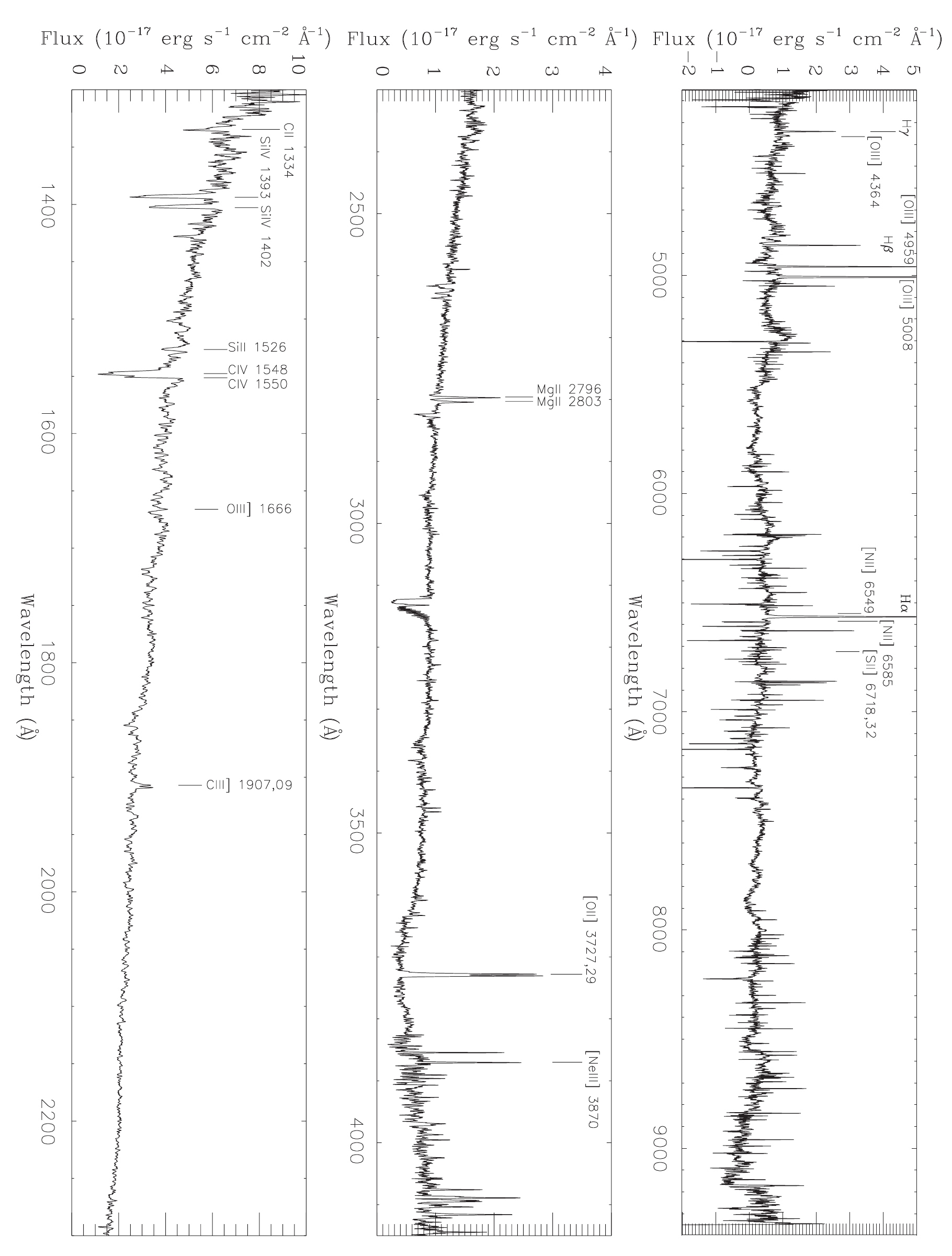}
\caption{X-Shooter spectrum of CSWA\,20 
with the most important emission and absorption lines labelled;
\textit{top panel}: NIR arm,  \textit{middle panel}: VIS arm,  
and \textit{bottom panel}: UVB arm.  
Wavelengths are given in the rest frame of the gravitationally lensed
galaxy at $z_{\rm em} = 1.4335$.
Each spectrum has been smoothed with a 5-pixel boxcar for presentation purposes.} 
\label{fig:spec}
\end{figure*}

Figure~\ref{fig:spec} shows the complete, 
flux-calibrated X-Shooter spectrum of CSWA\,20.  
The spectrum exhibits all the strong emission lines 
typical of star-forming galaxies, 
superimposed on a blue continuum.  
On the other hand, the UV absorption spectrum
is unusual, in that most interstellar lines are much
weaker than is normally the case in star-forming galaxies at these
redshifts \citep[e.g.][]{Steidel:2004, Erb:2010, Martin:2012, Kornei:2013}.
In this Section, we describe the main characteristics of the 
spectrum of CSWA\,20, focussing on the emission lines
in Section~\ref{sec:emission}, and on the absorption
lines in Section~\ref{sec:AbsLines}.


\begin{table}
\caption{Observed and de-reddenened line fluxes relative to \hb.$^{\rm a}$}
\label{tab:EmLines}
\begin{center}
\begin{tabular}{lcc}
\hline\hline
	&	$F_\lambda$			&	$I_\lambda$			\\
\sfoiii~$\lambda 1660.81$    &	0.04	$\pm$	0.04	&	0.11	$\pm$	0.11	\\ 
\sfoiii~$\lambda 1666.15$    &	0.10	$\pm$	0.04	&	0.26	$\pm$	0.13	\\
\fciii~$\lambda 1906.68$     &	0.12	$\pm$	0.01	&	0.28	$\pm$	0.08	\\
\sfciii~$\lambda 1908.73$    &	0.11	$\pm$	0.01	&	0.25	$\pm$	0.07	\\
\mgii~$\lambda 2796.36$	     &	0.22	$\pm$	0.01	&	0.33	$\pm$	0.05	\\
\mgii~$\lambda 2803.53$      &	0.15	$\pm$	0.01	&	0.22	$\pm$	0.03	\\
\foii~$\lambda 3727.10$	     &	0.47	$\pm$	0.02	&	0.55	$\pm$	0.04	\\
\foii~$\lambda 3729.86$	     &	0.52	$\pm$	0.03	&	0.62	$\pm$	0.05	\\
\fneiii~$\lambda 3870.16$    &	0.37	$\pm$	0.02	&	0.43	$\pm$	0.03	\\
\hg~$\lambda 	4341.69$     &	0.57	$\pm$	0.05	& 	0.61	$\pm$	0.06	\\
\foiii~$\lambda 4364.44$     &	0.05	$\pm$	0.03	&	0.05	$\pm$	0.03	\\
\hb~$\lambda 4862.69$        &	1.00	$\pm$	0.07	&	1.00	$\pm$	0.07	\\
\foiii~$\lambda 4960.30$     &	1.67	$\pm$	0.10	&	1.65	$\pm$	0.10	\\
\foiii~$\lambda 5008.24$     &	4.97	$\pm$	0.29	&	4.88	$\pm$	0.28	\\
\fnii~$\lambda 6549.85$	     &	0.02	$\pm$	0.04	&	0.02	$\pm$	0.03	\\
\ha~$\lambda 6564.61$ 	     &	3.32	$\pm$	0.20	&	2.78	$\pm$	0.24	\\
\fnii~$\lambda 6585.27$	     &	0.06	$\pm$	0.02	&	0.05	$\pm$	0.02	\\
\fsii~$\lambda 6718.29$	     &     0.06	$\pm$	0.03	&	0.05	$\pm$	0.03	\\
\fsii~$\lambda 6732.67$	     & 	0.12	$\pm$	0.02	&	0.10	$\pm$	0.02	\\
\\									
$F$(\hb)	& \multicolumn{2}{c}{		$(2.38	\pm	0.12) \times10^{-16}$\,erg~s$^{-1}$~cm$^{-2}$}	\\
$I$(\hb)	& \multicolumn{2}{c}{		$(4.21	\pm	0.34) \times10^{-16}$\,erg~s$^{-1}$~cm$^{-2}$}	\\
\\
$E(B-V)$	& \multicolumn{2}{c}{ 	$0.17	 \pm 	0.06$\,mag	}				\\
\hline
\end{tabular}
\end{center}
~~$^{\rm a}$ Vacuum wavelengths.
\end{table}

\subsection{Emission lines}
\label{sec:emission}


\begin{figure*}
\includegraphics[scale=0.675]{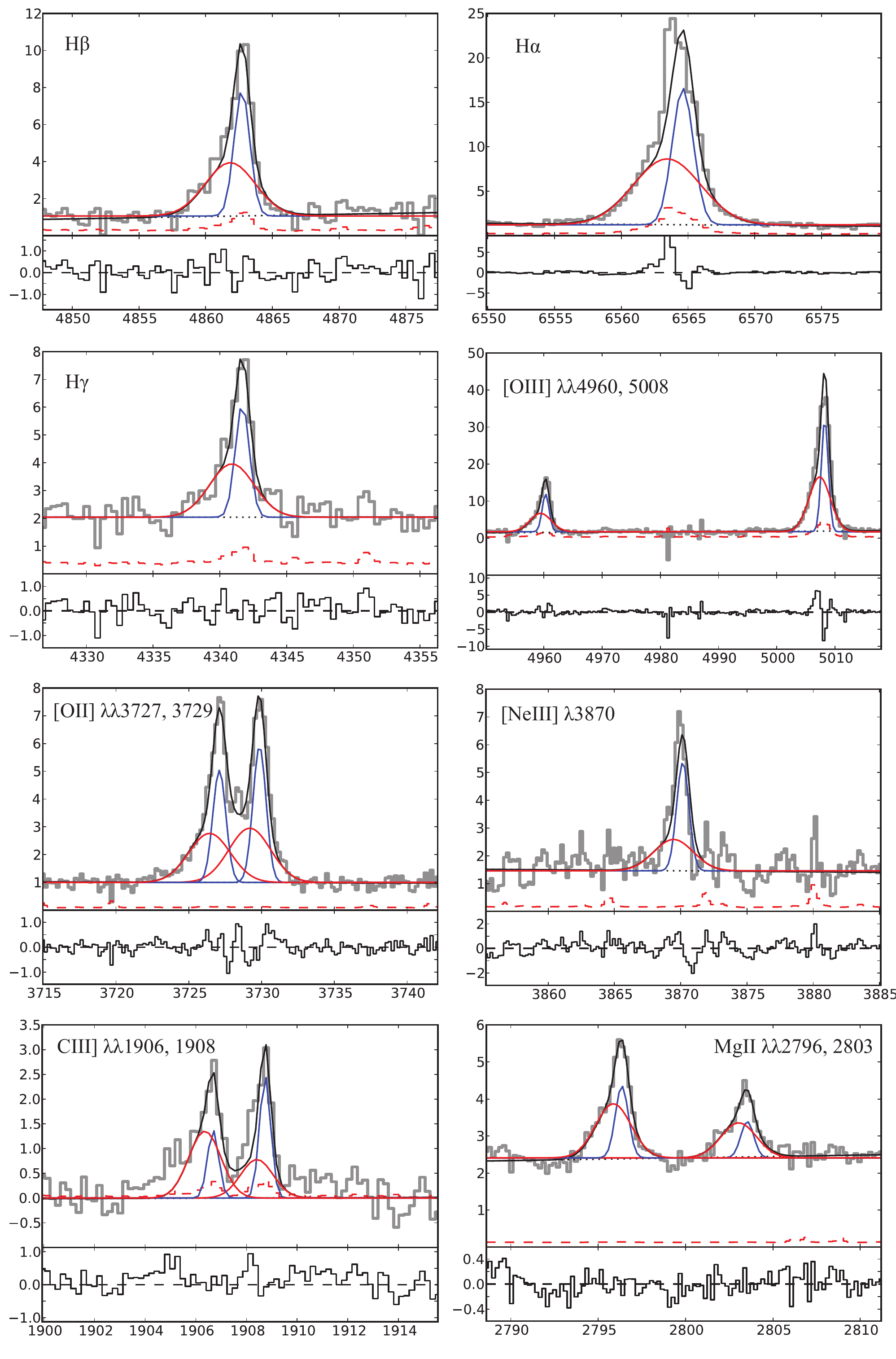}
\caption{High S/N emission lines observed within the X-shooter wavelength range. 
In each panel, the $x$-axis is  the rest-frame wavelength in \AA, 
and the $y$-axis is the observed flux in units of $10^{-17}$~erg\,s$^{-1}$\,cm$^{-2}$\,\AA$^{-1}$.  
In each plot, the grey histogram is the observed spectrum,
and the red dashed line is the $1 \sigma$ error spectrum.  
Overplotted on the data we show separately  
the best-fit profiles for the narrow (blue solid line), 
and broad (red solid line) emission components; 
their sum is shown with the black solid line.  
Relevant parameters of the model fit are 
given in Table~\ref{tab:params}, and the total line fluxes are
listed in Table~\ref{tab:EmLines}.
Underneath each emission line 
we show the difference between the data and the best-fit model.
}
\label{fig:lines}
\end{figure*}

In Figure~\ref{fig:lines} we have collected the strongest emission lines
observed in our spectrum. All of them exhibit asymmetric profiles,
with an extended blue wing. 
We have modelled these lines with two separate Gaussian components
(the minimum required for a satisfactory fit), to deduce their redshifts
and velocity dispersions, as well as the total integrated line flux of each
emission line. Results of the model fits are collected in Tables~\ref{tab:params}
and \ref{tab:EmLines}.


\begin{figure*}
\includegraphics[scale=0.6]{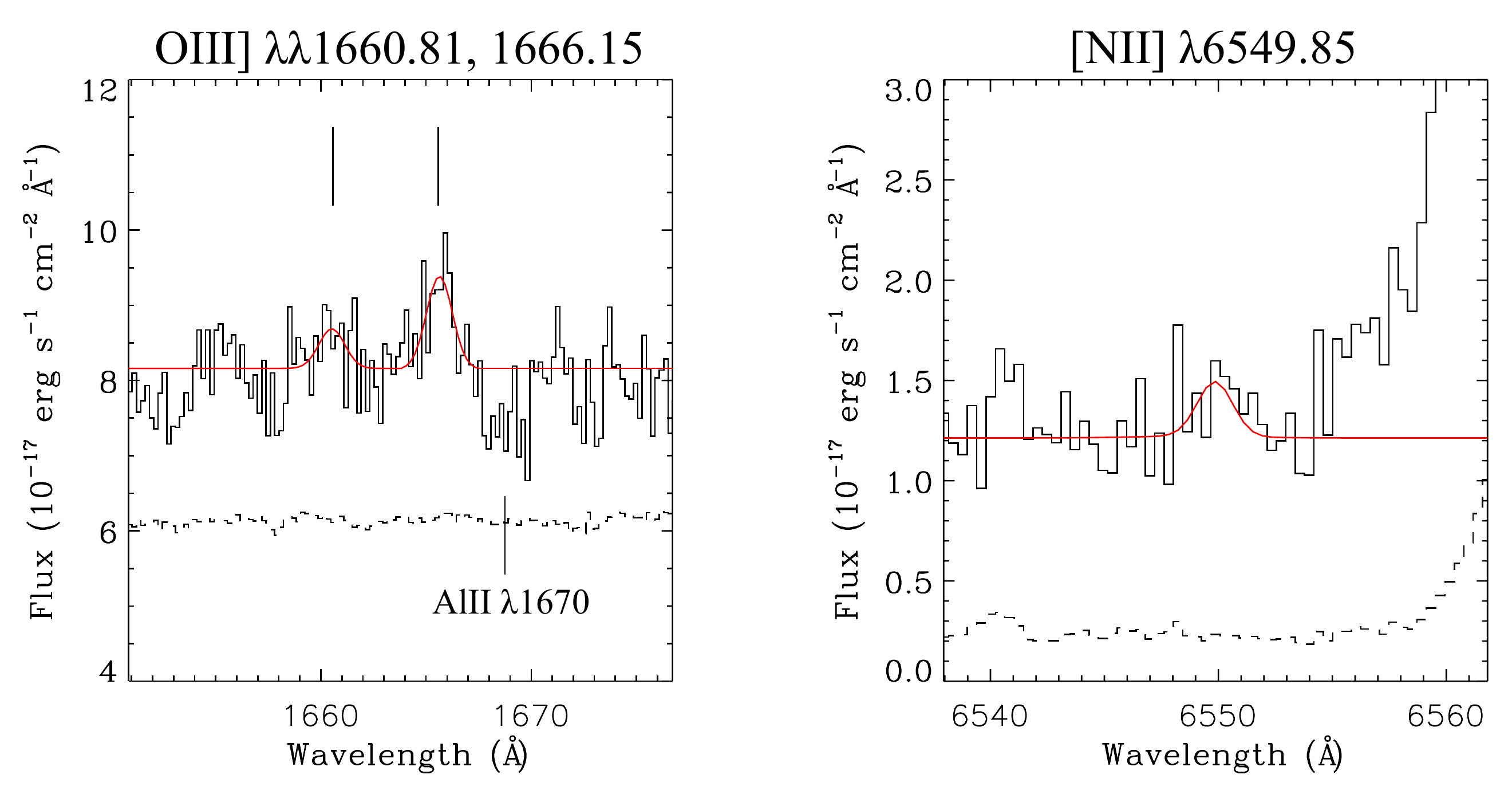}
\caption{Example of weak emission lines near, or below, the detection limit
of the X-shooter spectrum of CSWA\,20.
\sfoiii~$\lambda 1666.15$ is detected at the $\sim 3 \sigma$ level,
while \sfoiii~$\lambda 1660.81$ and 
\fnii~$\lambda 6549.85$ are comparable to the noise
(see Table~\ref{tab:EmLines}).
The data are the black histogram and the model fit is shown by the
red continuous line.
The dashed line is the error spectrum; in the case of the 
\sfoiii~doublet (left panel), we have added a 
constant factor of $5 \times 10^{-17}$\,erg~s$^{-1}$~cm$^{-2}$~\AA$^{-1}$
to the noise spectrum in order to bring it within the range of the plot.
Also marked on this plot is the interstellar \alii\ absorption line.}
\label{fig:weak_lines}
\end{figure*}

All of the emission lines shown in Figure~\ref{fig:lines} were fitted
simultaneously (with the exception of the \mgii\ doublet for
the reasons given below), to ensure consistent kinematic measurements,
(i.e., we required the redshifts and line widths of each 
emission component to be the same for all the lines fitted).
The model included four non-linear parameters (the redshifts $z$ and
velocity dispersions $\sigma$ of each velocity component)
that were determined by a Markov chain Monte Carlo (MCMC) method,
while the fluxes in each emission component were treated as linear parameters.
For each MCMC step we solved the bounded linear problem 
that determines the (non-zero) amplitude of each line 
given the current values for the non-linear parameters;
we explicitly accounted for the uncertainties 
from the linear inversion by attaching samples 
for the inference of the line amplitudes to the MCMC chain.

The \mgii~$\lambda\lambda2796, 2803$ doublet lines
were fitted separately from the other strong emission lines 
because, being resonance lines, they are potentially a mix
of absorption and emission. Indeed, in galaxies hosting
large-scale outflows of the interstellar medium (ISM), 
these lines can exhibit P-Cygni profiles, akin to those
more commonly seen in Ly$\alpha$ \citep[][]{Rubin:2010, Kornei:2013}.
The finding that in CSWA\,20 the \mgii\ doublet is mostly in
emission (see bottom right panel of Figure~\ref{fig:lines})
is likely to be a consequence of the unusually low optical depth
in this galaxy
of interstellar absorption lines from neutral, or mildly ionised,
gas, as discussed below (Section~\ref{sec:AbsLines}).
As can be seen from Table~\ref{tab:params}
and Figure~\ref{fig:lines}, 
the \mgii\ lines also consist of two components,
with velocity dispersions which are broadly consistent
with those determined for the nebular lines within the errors.
On the other hand, the velocities of the two components 
are higher than those of the nebular lines by $\sim 35$\,km~s$^{-1}$,
probably as a result of resonant scattering.

In addition to the emission lines collected in Figure~\ref{fig:lines}, 
the X-shooter spectrum covers a number of other weak emission lines
which are near, or below, the detection limit. 
Three examples are reproduced in Figure~\ref{fig:weak_lines}.
The fluxes of these weak features are included in Table~\ref{tab:EmLines};
they were measured by
fitting to the data theoretical profiles with the model parameters  
determined for the high S/N emission lines, as listed in Table~\ref{tab:params}.

\subsubsection{Reddening and star-formation rate}
\label{sec:SFR}

From Table~\ref{tab:EmLines} it can be seen that we measure 
$F$(\ha)/$F$(\hb)$=3.32 \pm 0.20$. 
Adopting the case B recombination ratio
$F$(\ha)/$F$(\hb)$=2.78$ from \citet{Brocklehurst:1971},
appropriate to gas with $T = 17\,000$\,K and $n{\rm (e)} = 300$\,cm$^{-3}$
as derived below for CSWA\,20
(see Section~\ref{sec:TeNe}),
and the LMC extinction curve \citep{Fitzpatrick:1999},
we find that the Balmer decrement implies 
a reddening $E(B-V) = 0.17 \pm 0.06$\,mag.\footnote{This value
is higher than that reported by \citet{Pettini:2010}, although
still consistent within the errors. We consider the present
estimate more reliable, given the higher S/N ratio
of the spectrum analysed here.}


\begin{figure*}
\includegraphics[scale=0.33, angle=-90]{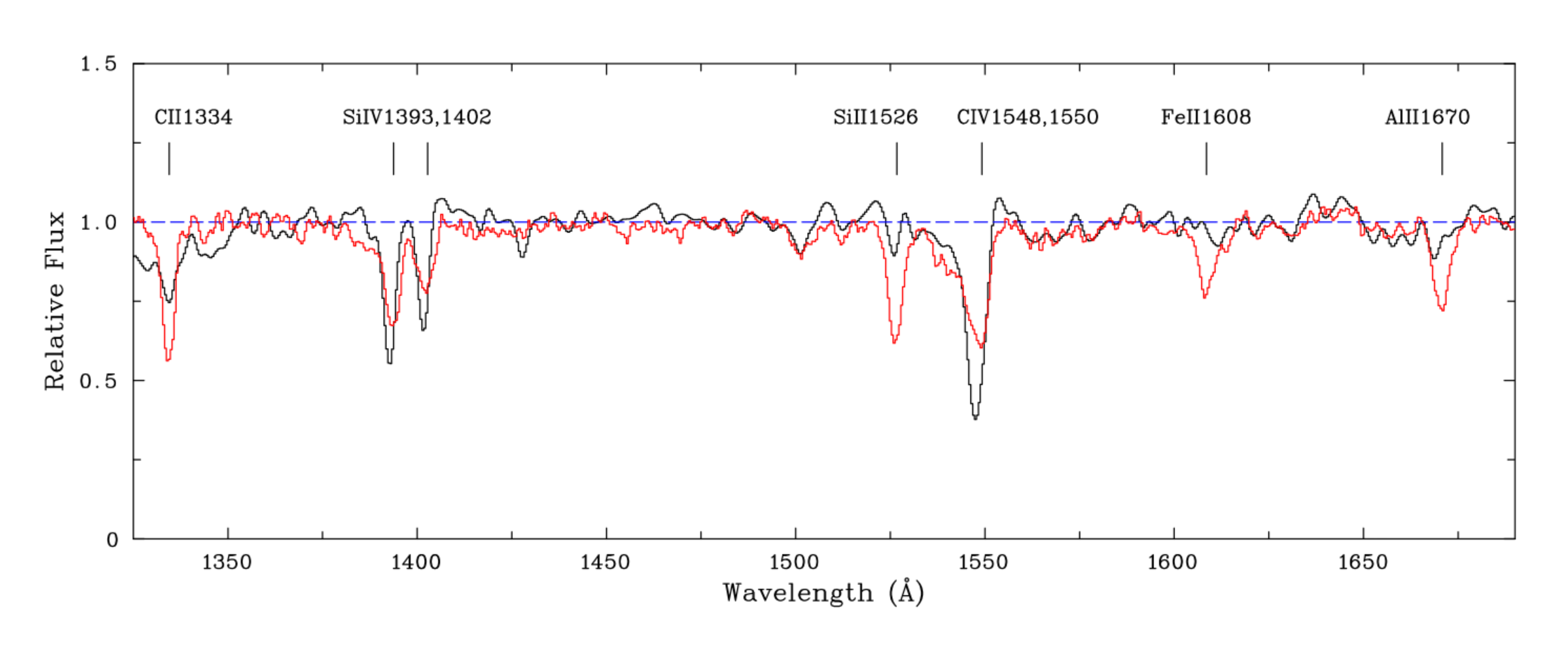}
\caption{Comparison of portion of the normalised spectrum of CSWA\,20
(black histogram) with the median spectrum of $\sim 380$ `BM' 
star-forming galaxies at $z \simeq 1.4$--2.0 
from the surveys by \citet{Steidel:2004}, shown in red.
Interstellar lines of interest are indicated. The X-shooter spectrum 
of CSWA\,20 has been smoothed and rebinned to match the $\sim 2.5$\,\AA\
resolution of the BM composite.} 
\label{fig:BMcomp}
\end{figure*}


\begin{table*}
\begin{center}
\caption{Interstellar Absorption Lines.}
\label{tab:AbsLines}
\begin{tabular}{lcccccc}
\hline
\hline
Line ID	&	$\lambda_{\rm lab}^{\rm a}$ (\AA)	&	$f$-value$^{\rm a}$	&	$\lambda_{\rm c}^{\rm b}$ (\AA)	&	$v^{\rm c}$ (km s$^{-1}$)	&	$\Delta v^{\rm d}$	&	$W_0^{\rm e}$ (\AA)			\\
\hline															
\hline															
\cii~$\lambda 1334$	         &	1334.5323	&	0.1278	&	1334.17	&	$-81$	&	$-540$ to $+270$	   &	$0.63 	\pm 	0.07$	\\
\civ~$\lambda 1548$	&	1548.2041	&	0.1899	&	...	&	...	&	$-740$ to $+190$	             &	$3.44 	\pm	0.07^{\rm f}$	\\
\civ~$\lambda 1550$	&	1550.7812	&	0.09475	&	...	&	...	&	$-740$ to $+190$	             &	$3.44 	\pm	0.07^{\rm f}$	\\
\sIii~$\lambda 1526$	&	1526.7070	&	0.133	&	1526.04	&	$-131$	&	$-540$ to $+270$	    &	$0.52 	\pm	0.07$	\\
\sIiv~$\lambda 1393$	&	1393.7602	&	0.513	&	1392.61	&	$-248$	&	$-740$ to $+190$   &	$1.78 	\pm	0.06$	\\
\sIiv~$\lambda 1402$	&	1402.7729	&	0.254	&	1401.57	&	$-257$	&	$-740$ to $+190$   &	$1.48 	\pm	0.06$\\
\alii~$\lambda 1670$	&	1670.7886	&	1.74	&	1669.54	&	$-224$	&	$-540$ to $+270$	   &	$0.36 	\pm	0.07$\\
\aliii~$\lambda 1854$	&	1854.7184	&	0.559	&	1853.30	&	$-229$	&	$-740$ to $+190$   &	$0.65 	\pm	0.08$\\
\aliii~$\lambda 1862$	&	1862.7910	&	0.278	&	1860.75	&	$-329$	&	$-740$ to $+190$   &	$0.42 	\pm	0.08$\\
\hline
\end{tabular}
\end{center}
\begin{description}
\item \textsc{Notes:}
\item $^{\rm a}$ Rest vacuum wavelengths and $f$-values from \citet{Morton:2003}.
\item $^{\rm b}$ Centroid rest-frame wavelength.
\item $^{\rm c}$ Velocity relative to systemic, at $z_{\rm em}$=1.433485.
\item $^{\rm d}$ Velocity range for equivalent width measurements relative to $z_{\rm em}$=1.433485.
\item $^{\rm e}$ Rest-frame equivalent width and $1\sigma$ error.
\item $^{\rm f}$ Refers to \civ~$\lambda$1549 doublet.
\end{description}
\end{table*}%

With the adoption of the conventional $\Omega_{\rm M} = 0.3$, $\Omega_{\Lambda} = 0.7$,
$h = 0.7$ cosmological parameters, the reddening-corrected H$\alpha$ flux
we measure in CSWA\,20 (Table~\ref{tab:EmLines}), corresponds to  
a luminosity 
$L{\rm (H}\alpha) = (1.49 \pm 0.13)  \times10^{43}$\,erg~s$^{-1}$.
This value can be converted into a star formation rate:
\begin{equation}
{\rm SFR(H\alpha)} = 7.9 \times 10^{-42} L({\rm H}\alpha) \times \frac{1}{1.8} \times \frac{1}{11.5}
= (5.7 \pm 0.5) \, {\rm M}_\odot~{\rm yr}^{-1}
\label{eq:SFR1}
\end{equation}
where the first factor on the right-hand side of the equation is
Kennicutt's (1998) calibration appropriate to a Salpeter 
initial mass function, corrected for the more realistic 
turn-over at low masses proposed by
\citet{Chabrier:2003} (amounting to a factor of 1.8),
and for the 11.5 magnification factor returned by our lensing
model. 
(Note that an additional systematic error of 10\% applies 
to all our estimates of line luminosity and SFR from the 
10\% uncertainty in the magnification factor -- see Section~\ref{sec:LensMod}).
The value of SFR in equation~(\ref{eq:SFR1}) is $\sim 3.7$
times smaller than that deduced by \citet{Pettini:2010},
mainly as a result of the better modelling of the 
lens system performed here.

The star formation rate can also be deduced from the luminosity
of the UV continuum produced by OB stars.
From the spectrum in Figure~\ref{fig:spec}, we
measure 
$F_\lambda (1500) = (4.7 \pm 0.5) \times 10^{-17}$\,erg~s$^{-1}$~cm$^{-2}$~\AA$^{-1}$ 
by averaging over a 40\,\AA\ window centred at 1500\,\AA\
away from obvious emission and absorption lines.
(The error quoted is from the $\sim 10$\% uncertainty in our
flux calibration---see Section~\ref{sec:DataRed}).
In our cosmology, this corresponds to a continuum luminosity density at 1500\,\AA:
$L_\nu = (9.2 \pm 4.4)  \times 10^{29}$\,erg~s$^{-1}$~Hz$^{-1}$,
after correcting for the attenuation at 1500\,\AA\
according to \citet{Calzetti:2000}.\footnote{
 We have adopted the starburst attenuation curve and scaling by \citet{Calzetti:2000}: $E(B-V)_{\rm stars} \simeq 0.44\, E(B-V)_{\rm nebular}$ which fits the UV continuum slope of CSWA\,20.}
(The error now includes the uncertainty in $E(B-V)$,
added in quadrature). In turn, this leads to:
 \begin{equation}
{\rm SFR(UV)} = 1.4 \times 10^{-28} L_\nu{\rm (UV)}\times \frac{1}{1.8} \times \frac{1}{11.5}
=(6 \pm 3) \, {\rm M}_\odot~{\rm yr}^{-1}
\label{eq:SFR2}
\end{equation}
following the same reasoning as in equation~(\ref{eq:SFR1}).

The larger error in SFR(UV) compared to SFR(H$\alpha$)
arises from the larger uncertainty in the attenuation at
1500\,\AA\ compared to the attenuation at H$\alpha$.
The two  estimates of the SFR are in good agreement.  It is not unexpected, however, for the two estimates to differ by 
a factor of $\sim 2$, given the many assumptions involved 
(including that the UV and H$\alpha$ luminosities
track each other, which is only the case in the 
ideal scenario of continuous star formation at a
constant rate). Differences by factors of $\sim 2$
between SFR(H$\alpha$) and SFR(UV) are 
commonly found in such estimates at high $z$
\citep[e.g.][]{Erb:2006c}.


\begin{figure*}
\includegraphics[scale=0.65,angle=90]{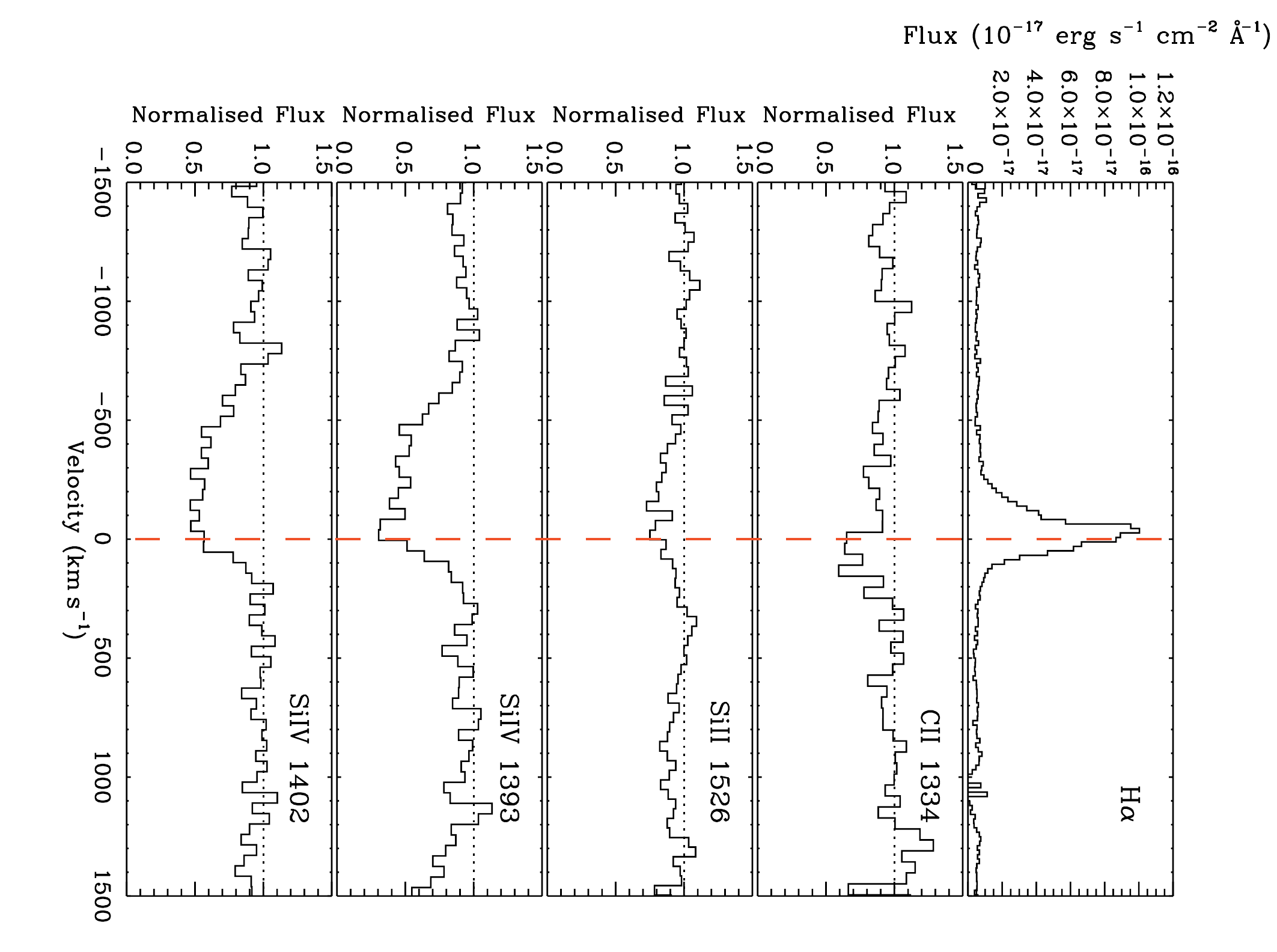}
\caption{Normalised profiles of selected interstellar absorption lines in CSWA\,20, 
compared to the \ha\, emission line (\textit{top panel}). 
The $x$-axis gives  velocity  relative to $z_{\rm sys}=1.433485$, 
as derived from the nebular emission lines (Section~\ref{sec:emission}).} 
\label{fig:abs}
\end{figure*}

\subsection{Absorption lines}
\label{sec:AbsLines}

It may be appreciated from inspection of Figure~\ref{fig:spec}
that very few interstellar absorption lines seem to be present
in the UV spectrum of CSWA\,20.
The main reason is the unusual weakness of absorption lines
from neutrals and the first ions that are the main ionisation
stages of the corresponding elements in H\,{\sc i} gas,
readily evident from Figure~\ref{fig:BMcomp}.
Table~\ref{tab:AbsLines} lists the absorption lines detected,
together with their rest-frame equivalent widths, $W_0$, and velocity range, 
$\Delta v$, over which  the absorption takes place. 
With $W_0 \simeq 0.5$--0.6\,\AA, \cii~$\lambda 1334$ 
and \sIii~$\lambda 1526$ are  $\gtrsim 3$ times
weaker than in the composites of star-forming galaxies at $z \simeq 2$--3
assembled by \citet{Shapley:2003} and \citet{Erb:2010}.
\oi~$\lambda 1302$ is barely detected (although this wavelength is strongly affected by noise),
nor do we see  absorption from \feii\, even in the strongest lines at 2344, 2382 and 2600\,\AA.

On the other hand, absorption lines from highly ionised gas,
such as \sIiv\ and \civ, appear to have similar, or even greater,
strengths in CSWA\,20 than in the above composites.
Evidently, the interstellar medium of this galaxy, as seen
from our line of sight, is mostly ionised, or the neutral gas
has a significant lower covering factor than the ionised gas.
In this respect, CSWA\,20 is reminiscent of the young,
unreddened, low-metallicity galaxy Q2343-BX418 studied
by \citet{Erb:2010}, although its UV emission lines are
not as strong, especially \heii~$\lambda 1640$ which usually 
denotes the presence of a population of Wolf-Rayet stars.
CSWA\,20 may well be a Ly$\alpha$ emitter although,
at its relatively low redshift, observations from space would
be required to verify that this is indeed the case.

\subsubsection{Kinematics of the absorbing gas}

The resolving power of X-shooter is sufficient to resolve the 
velocity profiles of the interstellar lines, and thereby probe
the kinematics of the interstellar medium of CSWA\,20.
A selection of lines is reproduced in Figure~\ref{fig:abs} 
on a velocity scale centred on the redshift of the narrow
component of the emission lines, $z_{\rm em} = 1.433485$,
which we take to be the systemic redshift, $z_{\rm sys}$, of the lensed galaxy.
Also shown for comparison  is the profile of the 
H$\alpha$ emission line (top panel).

Absorption lines from highly ionised gas 
(\sIiv\ and \aliii) are found to span nearly
$1000$\,km~s$^{-1}$, from
$\sim +200$\,km~s$^{-1}$ to 
$\sim -750$\,km~s$^{-1}$
relative to $z_{\rm sys}$.
The \civ\,$\lambda 1548, 1550$  lines are difficult 
to separate from each other and from their
stellar counterparts, but they are likely to extend  
over a similar $\Delta v$.
Weak absorption from the first ions
is also seen over a substantial 
fraction of this velocity range. 

In all cases, it is quite possible that gas may 
be present at higher velocities than can be measured
from our spectrum, as we now explain. 
It is evident from Figure~\ref{fig:abs}
that the absorbing gas does not fully cover
the source of the UV continuum, because
the \sIiv~$\lambda 1393, 1402$ lines 
exhibit comparable \textit{apparent} optical depths
throughout the velocity range over which
absorption is detected, even though their $f$-values
are in the ratio 2:1.  Thus, 
\sIiv\ absorption must in reality be saturated 
throughout, and its covering factor of the background
OB stars decrease with increasing 
negative velocity, as discussed by \citet{Steidel:2010}.
In this interpretation, the blue limit of the 
absorption lines is reached when the 
covering factor, rather than the ion column density, 
becomes too small for absorption to be detected 
at the S/N ratio of the present data.

In all cases where it could be measured,
the line centroids are blueshifted relative
to $z_{\rm sys}$ (fifth column of Table~\ref{tab:AbsLines}). 
This is a common occurrence
in high redshifts star-forming galaxies, 
which was recognised soon after their first spectra
were obtained \citep[e.g.][]{Pettini:2001, Shapley:2003}
and has since been quantified extensively
\citep[][and references therein]{Steidel:2010}.
The generally accepted interpretation
\citep[e.g.][]{Heckman:2002}  is that
such blueshifts reflect large-scale outflows 
of the interstellar medium of galaxies with high 
star-formation rate surface densities, driven
by the deposition of energy and momentum by the 
starburst phenomenon. Such `superwinds' have 
important implications for the regulation
of star formation and the dispersal of the 
products of stellar nucleosynthesis in the 
intergalactic medium. 

What is remarkable about the extent of the absorption
in CSWA\,20 is that $\Delta v$ is comparable to the 
values measured in the few other gravitationally
lensed star-forming galaxies at high $z$ where this measurement could be made, 
\textit{even though its star-formation rate
is up to $\sim 50$ times lower than
in the previously studied cases}. 
Most of the gravitationally lensed galaxies scrutinised
so far are intrinsically luminous ($L \gtrsim L^\ast$)
with star-formation rates
SFR\,$\simeq 50, 100, 100$, and 270\,$M_\odot$~yr$^{-1}$ in
MS~1512--cB58 \citep{Pettini:2000}, 
the `Cosmic Horseshoe' \citep[a.k.a. CSWA\,1, J1148+1930;][]{Quider:2009},
the `Cosmic Eye' \citep[J2135$-$0101;][]{Quider:2010}, and
the `Eight o'clock Arc' \citep[J0022$+$1431;][]{Dessauges:2010} respectively.
In contrast, for CSWA\,20 we deduced 
SFR\,$\simeq 6$--7\,\Msol~yr$^{-1}$ (Section~\ref{sec:SFR}).

Thus, on the basis of the very limited sample available at present,
there does not appear to be a strong correlation between 
the star formation rate of the starburst 
and the maximum velocity of the outflowing ISM, 
$v_{\rm max}$. On the other hand, 
the possibility raised above that
the measured value of  $v_{\rm max}$
may be more indicative of a very low covering fraction
rather than of an absence of gas at such velocities,
may well mask a correlation between SFR and  $v_{\rm max}$,
if one exists.

We also note in this context that \citet{Martin:2012}
did find a correlation between SFR and $v_{\rm max}$
measured from Fe\,{\sc ii} and Mg\,{\sc ii} absorption
lines in galaxies at  $0.4 < z < 1.4$ (see their Figure~14).
However, the correlation is only evident when 
considering values of SFR spanning nearly four orders
of magnitude, from less than 0.1\,M$_\odot$~yr$^{-1}$
to $\sim 400$\,M$_\odot$~yr$^{-1}$.
For SFRs greater than a few M$_\odot$~yr$^{-1}$
the data considered by \citet{Martin:2012} 
exhibit a great deal of scatter and little, if any,
dependence of $v_{\rm max}$ on SFR, consistent
with our conclusions here.

\section{Chemical Abundances}
\label{sec:abund}

In this Section we analyse the emission lines
detected in the spectrum of CSWA\,20 using 
established procedures to derive the abundances of
selected elements in the \hii\ regions of the galaxy. 
Specifically, a
`direct' measurement of the abundance of ion X$^{\rm n}$ 
relative to H$^+$ is made based on the electron temperature  
(\elt), and  density (\eld) of the gas and the reddening-corrected line fluxes. 
Ionic abundances are then converted into elemental abundances after accounting 
for unseen stages of ionisation.  

\subsection{Temperature and density}
\label{sec:TeNe}
\elt\, is normally deduced from the ratio of an auroral line to a lower excitation line, 
the most common line ratio used for this purpose being 
\foiii\,$\lambda$5008/$\lambda$4364.  
Auroral lines are notoriously weak, and uncertainties in their measurement can give 
large errors in the derived chemical abundances.  
The high magnification of CSWA\,20 and the large wavelength range of X-shooter 
give us access to two auroral lines: \foiii\,$\lambda 4364$ and \sfoiii\,$\lambda 1666$.  
Unfortunately, \foiii\,$\lambda 4364$ is affected by sky residuals
and is `detected' at only the $\sim 1.5 \sigma$ level.
The lower sky background at blue and ultraviolet wavelengths 
allows a clearer view of \sfoiii\,$\lambda 1666$,
which is detected at the $\sim 3 \sigma$ level 
(see Figure~\ref{fig:weak_lines}, Table~\ref{tab:EmLines}).
Thus, we used the ratio $I$(\sfoiii\,$\lambda 1666$)/$I$(\foiii\,$\lambda 5008$)
to deduce the value of \elt.
For the electron density, \eld, we have at our disposal
two density-sensitive line ratios:
\foii\,$\lambda 3727/ \lambda 3729$,
and \fciii\,$\lambda 1906$/\sfciii\,$\lambda 1908$.
Values of \elt\ and \eld\ were computed 
using \textsc{iraf's} {\tt temden} task in the {\tt nebula} package, 
and following the iterative algorithm utilised by the {\tt zones} task.  
We find  \elt\,$= 17000 \pm 3300$\,K.\footnote{We note 
that had we used \foiii\,$\lambda 4364$ to deduce the electron
temperature, we would have deduced
\elt\,$= 11700^{+5700}_{-1700}$\,K, in agreement with 
the value deduced above within the large errors.}

The electron temperature deduced in CSWA\,20 is higher than
the canonical \elt\,$\simeq 10000$\,K typical of \hii\ regions
in the local Universe, probably reflecting the low metallicity
of this galaxy deduced below (Section~\ref{sec:O/H}). 
On the other hand,
the value found here is in line with the similarly high  
electron temperatures deduced in the few
cases reported so far where weak auroral lines have been 
detected in high-$z$ star-forming galaxies,
\elt\,$\simeq 13000$--25000\,K 
\citep[][]{Villar:2004, Yuan:2009, Erb:2010, Christensen:2012b}.
However, this is likely to be a selection effect, given that
lower temperatures (and higher metallicities)
would result in auroral lines too weak to be detected
with the sensitivities of most previous studies
\citep[e.g.][]{Rigby:2011}.

It is noteworthy that the \sfciii\ line ratio,
despite its large error, indicates a considerably higher
electron density than the \foii\ doublet:
\eld\,$=17100 \pm 6100$\,cm$^{-3}$
 and
\eld\,$= 276 \pm 156$\,cm$^{-3}$
respectively.
Evidently, these two species arise in different 
ionisation zones with significantly different densities.
There are indications that this may be the case in other
high-$z$ star-forming galaxies where both doublets
have been observed
\citep[e.g.][]{Bayliss:2014, Christensen:2012b, Quider:2009, Hainline:2009},
although the uncertainties in the line ratios are often
too large to reach firm conclusions. This is an issue
that deserves to be addressed with better observations
in future.

\begin{table}
\begin{center}
\caption{Ionic and elemental abundances 
derived from the emission line measurements given in Table~\ref{tab:EmLines}.
\elt\ and \eld\ values are also listed.}
\label{tab:direct_abund}
\begin{tabular}{lc}
\hline
\hline
\elt(\foiii)	                    &	$(17000 \pm 3300)$\,K \\
\elt(\foii)$^{\rm a}$	&	$(14680 \pm 860)$\,K     \\
\eld(\foii)	                    &	$(276     \pm 156)$\,cm$^{-3}$	     \\
\eld(\sfciii)	          &	$(17100  \pm 6100)$\,cm$^{-3}$ \\
	                              &				      \\
\op/\hp	                    &	$(1.1 \pm 0.3) \times 10^{-5}$ \\
\opp\hp	                    &	$(5.6 \pm 1.1) \times 10^{-5}$  \\
O/H	                              &	$(6.7 \pm 1.4) \times 10^{-5}$ \\
$12+\log{\rm (O/H)}$	&	$7.82	\pm	0.21$  	\\
${\rm [O/H]}^{\rm b}$      &	$-0.87 \pm 0.21$      \\
	                              &				                \\
\cpp/\hp	                    &	$ (4.9 \pm 3.1) \times 10^{-6}$	\\
ICF(C)	                    &	$1.2 \pm 0.4$                           	\\
C/H	                              &	$(5.9   \pm 4.1) \times 10^{-6}$	\\
$12+\log {\rm (C/H)}$	&	$6.77 \pm 0.31$                  	\\
$\log {\rm (C/O)}$         &      $-1.06 \pm 0.32$                           \\
${\rm [C/O]^b}$         &	 $-0.80 \pm 0.33$ 	\\
	                              &				\\
\np/\hp	                    &	$(4.1	\pm 2.1) \times 10^{-7}$ 	\\
ICF(N)	                    &	$ 5.7 \pm	1.8$ 	\\
N/H	                             &	$(2.3 \pm 1.4) \times 10^{-6}$	\\
$12+\log {\rm (N/H)}$	&	$6.37 \pm 0.26$	                   \\
$\log {\rm (N/O)}$         &     $-1.46 \pm 0.28$                            \\
${\rm [N/O]^b}$	&	$-0.60 \pm 0.29$	\\
	                             &				\\
\nepp/\hp	         &	$(1.1 \pm 0.3) \times 10^{-5}$	\\
ICF(Ne)	                   &	$1.1 \pm 0.3$	\\
Ne/H	                   &	$(1.2 \pm 0.5) \times 10^{-5}$	\\
$12+\log{\rm (Ne/H)}$	&	$7.09	 \pm 0.17$	                   \\
$\log {\rm (Ne/O)}$     &       $-0.74 \pm 0.17$                            \\
${\rm [Ne/O]^b}$	&	$ + 0.02 \pm 0.22$	\\
	                             &				\\
\Sp/\hp	&	$(2.4 \pm 0.5) \times 10^{-7}$	                    \\
\Spp/\hp$^{\rm c}$	&	$(2.2 \pm 0.8) \times 10^{-6}$  \\
ICF(S)	                    &	$1.25 \pm	0.4$ 	                              \\
S/H	                              &	$(3.1 \pm 1.6) \times 10^{-6}$	\\
$12+\log{\rm (S/H)}$	&	$6.48	\pm 0.22$	\\
$\log {\rm (S/O)} $        &     $-1.34 \pm 0.22$         \\
${\rm [S/O]^b}$	&	$+0.23 \pm 0.25$	\\
\hline
\end{tabular}
\begin{description}
$^{\rm a}$ Derived using the formula given by \citet{Izotov:2006}.\\
$^{\rm b}$  ${\rm [X/Y]} \equiv \log {\rm (X/Y)}_{\rm CSWA\,20} - \log {\rm (X/Y)}_\odot$,
where $\log {\rm (X/Y)}_\odot$ are the solar composition values recommended by \citet{Asplund:2009}. \\
$^{\rm c}$ Derived using the formula given by \citet{Barlow:1994}.
 \end{description}
\end{center}
\end{table}

\subsection{Chemical composition}
\label{sec:O/H}

Ionic abundances relative to \hp\ were derived for the \cpp, \np, \op, \opp, \nepp\ and \Sp\ ions 
from the dereddened fluxes of, respectively,  $\lambda 1908$; $\lambda 6585$;
$\lambda \lambda 3727, 3729$; $\lambda \lambda 4960, 5008$;
$\lambda 3870$; and $\lambda \lambda 6718, 6732$ emission lines
(Table~\ref{tab:EmLines}) using the {\tt zones} task in \textsc{iraf}
with the values of  \elt\ and \eld\ deduced above (Section~\ref{sec:TeNe}).
The oxygen abundance can be derived directly by
adding together the \opp /\hp\, and \op /\hp\, ratios.
For the other elements, it is necessary to include
ionisation correction factors (ICFs) 
to account for unseen ionisation stages; to this end, we adopted the ICFs proposed by \citet{Izotov:2006} 
for all elements apart from C, for which we adopted the ICF of \citet{Barlow:1994}.
To account for the unobserved S$^{2+}$ ion
we have adopted the empirical relationship between
\Spp/\Sp and \opp/\op found by
\citet{Barlow:1994} in planetary nebulae (their equation A38),
but it is questionable whether the same scaling
applies to \hii\ regions which are ionised by a 
softer spectrum than PNe.  For this reason, the correction for S is particularly uncertain.

Abundance determinations
and ICFs  are collected in Table~\ref{tab:direct_abund}.
We find $12+\log{\rm (O/H)} = 7.82 \pm 0.21$ in CSWA\,20,
or [O/H]\,$= -0.87 \pm 0.21$ (approximately between 1/5 and 1/10 solar) 
relative to the solar value  $12+\log{\rm (O/H)}_\odot = 8.69$ \citep{Asplund:2009}.
The oxygen abundance in  CSWA\,20 is comparable 
to those of the least chemically enriched 
galaxies at $z \sim 2$ \citep{Erb:2006a}, as found in
other high-$z$ galaxies in which 
auroral \foiii\  lines have been detected
(likely a selection effect for the reasons explained in Section~\ref{sec:TeNe}).

As can be seen from Table~\ref{tab:direct_abund},
Ne and S are  in their solar proportions
relative to O, as expected given that they are both
$\alpha$-capture elements.
The N/O ratio is close to the plateau at
$\log {\rm(N/O)} \simeq -1.5$ ($\sim 1/4$ of the solar ratio)
found in galaxies with 
$12  + \log{\rm (O/H)} \lesssim 8.0$
\citep[e.g.][and references therein]{Izotov:2006}.
This plateau is thought to correspond 
to the level of `primary' production of N,
synthesised from seed C and O 
produced by the star during He burning
\citep[see][]{Henry:2000, Pettini:2008}.
The C/O ratio is discussed separately below.


\begin{figure*}
\includegraphics[scale=0.6,angle=90]{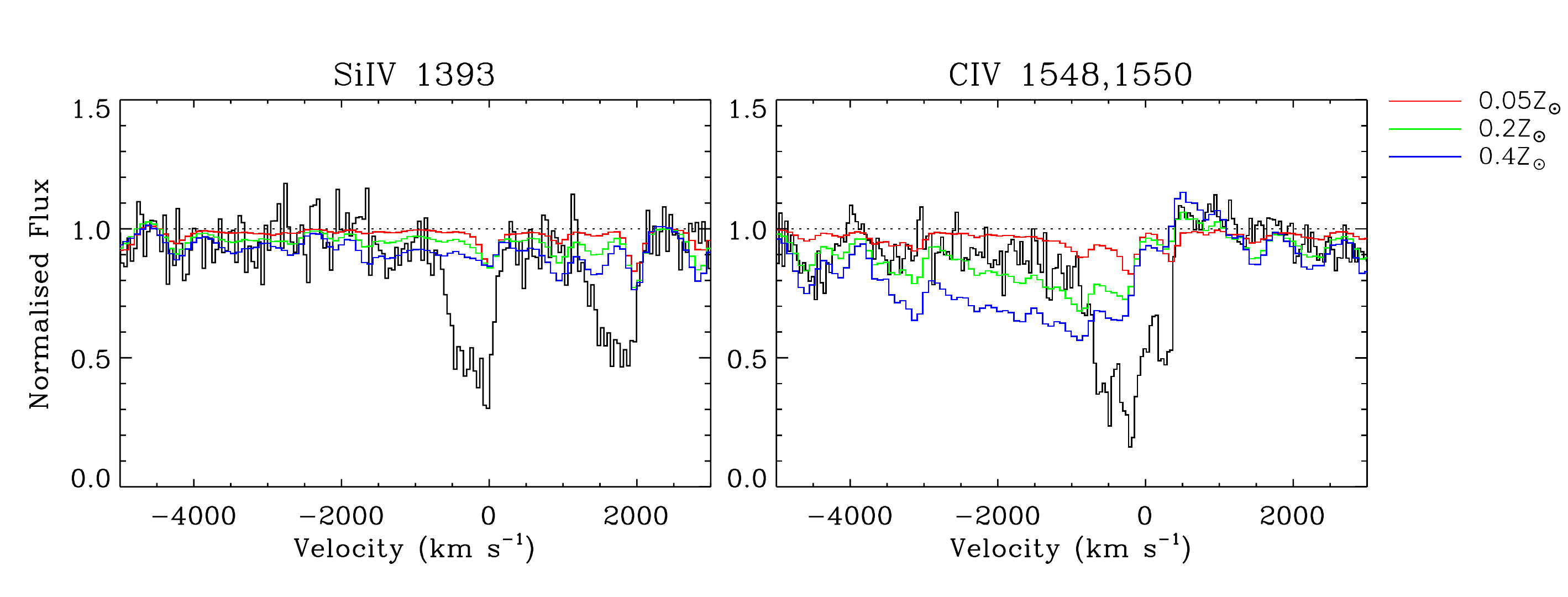}
\caption{Velocity profiles of high-ionisation lines arising from the winds of massive stars.
In each panel, the black histogram is the data for CSWA\,20, plotted on a 
velocity scale relative to $z_{\rm sys} = 1.433485$; the absorption features
seen in these portions of the spectrum are blends of stellar and interstellar 
absorption lines. Superimposed on the data are theoretical stellar
lines profiles for different metallicities computed with the
Starburst99 spectral synthesis code and convolved to the same resolution as the data.
In all three cases shown, we assumed continuous star formation,
100\,Myr age, and a Salpeter IMF between 1 and 100\,M$_\odot$.
Further details are given in Section~\ref{sec:WindLines}.} 
\label{fig:WindLines}
\end{figure*}

\subsubsection{C/O abundance ratio}
\label{sec:C/O}

The correction we have applied to convert
C$^{2+}$/H$^+$ to C/H
assumes that \cpp/\opp\,=\,C/O.
This may however not always be the case,
given that both \cp\ and \cpp\ have lower ionisation potentials
than \op\ and \opp.
This issue has recently been addressed by
\citet{Erb:2010} \citep[see also][]{Garnett:1995, Shapley:2003}
with the aid of Cloudy photoionisation modelling.
\citet{Erb:2010} found 
that \cpp/\opp\,$ \simeq$\,C/O only
for a relatively narrow range of values of the ionisation
parameter, $U$, customarily defined as the ratio of the densities
of ionising photons and particles.
For the case in hand, we estimated the value of $U$ 
using the iterative equations of \citet{Kewley:2002} 
together with the value of O/H derived for CSWA\,20 and the 
\foiii~$\lambda 5008$/\foii~$\lambda\lambda 3727, 3729$ ratio.  
Interpolating between the 0.1\,\Zsol\ and 0.2\,\Zsol\
equations of \citet{Kewley:2002} we find $\log U = -2.47 \pm 0.09$
(the error quoted includes both the uncertainties in the measurements
of the  line fluxes and the difference between the 
0.1\,\Zsol\ and 0.2\,\Zsol\ cases, added in quadrature).
Inspection of  Figure~11 of \citet{Erb:2010} confirms that,
in this range of values of  $U$, \cpp/\opp\ is indeed a valid 
approximation to C/O.

As can be seen from Table~\ref{tab:direct_abund}, 
in CSWA\,20 the carbon abundance relative to oxygen
is significantly lower than in the Sun: 
$\log ({\rm C/O})_{\rm CSWA\,20} = -1.06 \pm 0.32$,
between $\sim 1/3$ and $\sim 1/14$ of the solar value
$\log ({\rm C/O})_{\odot} = -0.26 \pm 0.07$ \citep{Asplund:2009}.
This value is lower than measured in nearby dwarf galaxies
of comparable oxygen  abundance
\citep[e.g.][]{Garnett:1995}, although still consistent
within the errors. 
\citet{Christensen:2012b} also deduced
$\log ({\rm C/O}) = -1.03 \pm 0.08$
and $-0.80 \pm 0.09$ in two high-$z$
gravitationally lensed galaxies 
with $12 + \log{\rm (O/H)} = 7.69 \pm 0.13$
and $7.76 \pm 0.03$ respectively, suggesting that
an underabundance of C relative to O by one order of magnitude
may not be unusual in such metal-poor environments.

There is continuing interest in measuring the 
C/O ratio at low metallicities
\citep[e.g.][]{Akerman:2004, Cooke:2011}
as a probe of stellar nucleosynthesis in this regime.
The well-established decrease in C/O with decreasing 
O/H has been interpreted as evidence for metallicity-dependent
yields of massive stars \citep{Akerman:2004} and/or 
delayed release of carbon into the ISM by stars of intermediate
and low mass \citep[e.g.][]{Chiappini:2003, Carigi:2005}.
In \citet{Akerman:2004} and \citet{Erb:2010} it was pointed out
that the similarity in the run of C/O vs. O/H between
different locations and epochs in the Universe 
favoured the first explanation.
Here we note that the lowest values of C/O measured in actively
star-forming galaxies at high redshifts seem to be lower
than the lowest measures of this ratio in Galactic halo stars
\citep{Akerman:2004} by $\sim 0.4$\,dex.  
Quite possibly, such a difference may simply
reflect systematic offsets in the derivation of
element abundances from stellar absorption spectra 
and \hii\ region emission spectra respectively.
On the other hand, if the difference is real, it would suggest
that both metallicity dependent yields and
evolutionary effects (i.e. the past history of star 
formation), may determine the C/O ratio one
would measure at a given oxygen abundance.
In this scenario, the lowest values of C/O 
found in high-$z$ star-forming galaxies, such as CSWA\,20,
would reflect a particularly rapid progress of
chemical evolution, with insufficient
time for lower mass stars to evolve and 
contribute their carbon to the ISM.

\section{Metallicity Indicators}
\label{sec:indicators}

As explained in the Introduction, one of the motivations for pursuing additional
observations of CSWA\,20 was the possibility of comparing different estimates
of the metallicity thanks to the good match between the redshift of the
galaxy and the wavelengths covered with the X-shooter spectrograph.
The discussion in Section~\ref{sec:emission} demonstrates that,
even with the aid of gravitational lensing, the detection of temperature
sensitive auroral lines is often beyond the reach of current astronomical
instrumentation.  This has led to the development
of approximate indicators of the oxygen abundance which
use only strong nebular emission lines. We discuss two
of these methods in Sections~\ref{sec:R23} and \ref{sec:N2} respectively.
In Sections~\ref{sec:WindLines} and \ref{sec:1978index}
we consider two alternative estimates of the \textit{stellar}
metallicity based on the strengths of, respectively,  the 
\civ\ (and \sIiv) lines produced in the winds of the most luminous
early-type stars, and a broad blend of Fe\,{\sc iii}  absorption lines
near 1978\,\AA\ from the photospheres of B-type stars.

\subsection{R$_{23}$ Index}
\label{sec:R23}
The $R_{23}$ index first proposed by \citet{Pagel:1979}
is the most widely used approximate indicator of the oxygen 
abundance in star-forming galaxies being based on the 
ratios of what are typically the strongest emission lines
from \hii\ regions  at visible wavelengths:  
$R_{23}\equiv[I(3727)+I(3729)+I(4960)+I(5008)]/I$(\hb).
Its main drawback, in principle, is that the relationship
between $R_{23}$ and O/H is double-valued, so that other
line ratios may be required to break the degeneracy. 
Uncertainties in the reddening may also be an issue,
given that the lines in question are spread 
over $\sim 1300$\,\AA\ of the optical spectrum. 
Regarding its application to high-redshift galaxies, however,
the main difficulty encountered with ground-based
observations is that all five emission lines fall
within near-infrared atmospheric transmission windows 
only for a limited set of redshifts.


\begin{figure*}
\hspace{-0.25cm}\includegraphics[scale=0.375,angle=90]{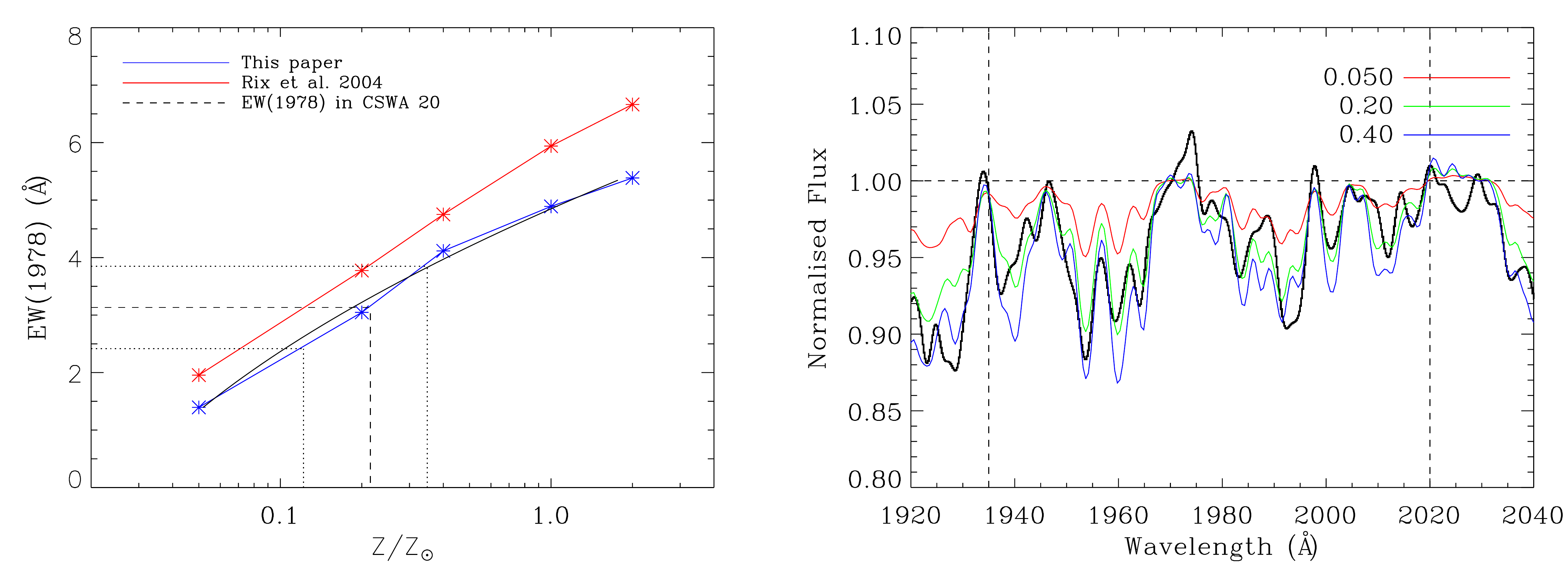}
\caption{\textit{Left:} Comparison of the old and new calibrations of the 
1978 index, as indicated. 
The black line is the functional fit to the new
calibration given in equation~(\ref{eq:1978}).
Dashed and dotted lines show the value of the 1978 index
measured in CSWA\,20.~
\textit{Right:} Portion of the spectrum of CSWA\,20
between 1920 and 2040\,\AA\ 
(smoothed to FWHM\,$= 2.5$\,\AA, black line), together
with synthetic Starburst99 spectra for three metallicities
(coloured lines). The integration limits for the 1978 index
are indicated by vertical dash lines.
In both panels, all models are for 100\,Myr old, continuous
star formation with a Salpeter IMF.
See Section~\ref{sec:1978index} for further details.
}
\label{fig:1978index}
\end{figure*}

At the redshift of CSWA\,20, $z_{\rm sys} = 1.433$, 
all the emission lines required to measure the $R_{23}$ index 
are free from strong water vapour absorption.
In deducing O/H from the unreddened line fluxes  in
Table~\ref{tab:EmLines}, we made 
use of the analytical expressions by \citet{McGaugh:1991} 
as given by \citet{Kobulnicky:1999}. 
These expression take into account the effect of
the ionization parameter on the relationship between
$R_{23}$ and O/H by including a term which depends on
the ratio of the \foii\ and \foiii\ lines,
$O_{32} \equiv[I(4960)+I(5008)]/[I(3727)+I(3729)]$.
We find 1$2+\log{\rm (O/H)} =7.96 \pm 0.11$ 
and $8.59 \pm 0.06$
for the low- and high-metallicity branches of the
$R_{23}$ index respectively.

\subsection{$N2$ Index}
\label{sec:N2}

The $N2$ index \citep[][]{Denicolo:2002, Pettini:2004}
has the advantage of involving only two emission lines 
at closely spaced wavelengths [$N2\equiv I(6585)/I$(\ha)],
thereby avoiding the need to correct for extinction.
On the other hand, it is only an approximate measure of 
O/H, with a precision of a factor of $\sim 2.5$ at the
95\% confidence level \citep{Pettini:2004}. Furthermore,
the index saturates near solar metallicity. 
Nevertheless, it has had widespread application 
for galaxies at redshifts $z \lesssim 2.5$ (at higher redshifts
the lines move beyond the atmospheric $K$-band window),
thanks to its simplicity and convenience.

With the calibration by \citet{Pettini:2004},
the fluxes in Table~\ref{tab:EmLines} imply
$12+\log{\rm (O/H)} = 7.91 \pm 0.09$
(where the uncertainty includes only the errors
in the line fluxes).  The same value is obtained 
with the more recent calibration of the $N2$ index by 
\citet{Marino:2013}. Thus, the $N2$ ratio identifies 
the lower-branch solution of the $R_{23}$ method
as the one applicable to CSWA\,20.

\subsection{Stellar wind lines}
\label{sec:WindLines}

The most prominent features in the rest-frame UV spectrum of CSWA\,20 
are absorption lines of \civ~$\lambda\lambda 1548, 1550$
and \sIiv~$\lambda\lambda 1393, 1402$ 
(see Figures~\ref{fig:spec} and \ref{fig:BMcomp}).
These lines are blends of interstellar absorption and P-Cygni 
emission-absorption profiles from the winds of the most 
luminous OB stars in the galaxy. 
Over the last fifteen years, Claus Leitherer
and collaborators have explored extensively 
how,  in integrated galaxy spectra, the strength of the wind 
component of these lines depends 
on the age, initial mass function 
and metallicity of the stellar population
\citep[e.g.][]{Leitherer:2011}.

In the limiting case of continuous star formation at a constant
rate with a universal IMF, the wind lines can be used as a
metallicity indicator \citep[e.g.][]{Mehlert:2002, Quider:2010}. 
Here, the `metallicity' refers mostly to the
abundances of C, N, O, and Fe,  
whose ions are the main contributors to the opacity 
driving the stellar winds \citep{Vink:2001}.
The attraction of this method is that it can applied 
at higher redshifts than the nebular emission line diagnostics
(which, from the ground, are restricted to galaxies at $z \lesssim 3.8$);
its limitations are the difficulty in separating stellar and interstellar
absorption \citep{Crowther:2006b}, and the relatively coarse
grid of stellar models currently available \citep{Leitherer:2010}.

In Figure~\ref{fig:WindLines}, we compare portions
of the normalised spectrum
of CSWA\,20\footnote{The spectrum was normalised 
using the continuum windows recommended by \citet{Rix:2004},
with minor modifications appropriate the the X-shooter data.}
with synthetic profiles generated by
Starburst99 with the WM-basic library of theoretical
stellar spectra \citep{Leitherer:2010}. 
In generating the Starburst99 spectra, we assumed a 
Salpeter slope for the IMF between 1 and 100\,M$_\odot$,
continuous star formation at a constant rate over 100\,Myr
(at this age the UV spectrum has stabilised and 
no longer depends on the age of the starburst),
and considered three metallicities of the OB stars:
0.05, 0.2 and 0.4\,\Zsol.  
The Starburst99 outputs were
convolved to the resolution of our X-shooter spectrum
(FWHM\,$\sim 0.9$\,\AA).

The left panel of Figure~\ref{fig:WindLines} shows that,
at these metallicities, the \sIiv\ stellar lines are very weak
and of limited diagnostic value; the absorption we see
is mostly interstellar. In the right panel, on the other hand,
it is possible to distinguish the weak emission and broad, 
shallow blue wing of the stellar P-Cygni profile extending
out to $\sim -4000$\,km~s$^{-1}$ from the narrower 
interstellar \civ\ absorption. The Starburst99 model
with $Z = 0.4$\,\Zsol\ clearly overpredicts both emission
and absorption components of the P-Cygni profile,
whose overall strength is intermediate between those of the 
0.2\,\Zsol\ and 0.05\,\Zsol\ models. We conclude that 
the metallicity of the most luminous OB stars in 
CSWA\,20 is $Z_{\rm OB} \sim 0.1^{+0.1}_{-0.05}$\,\Zsol.


\begin{figure*}
\includegraphics[scale=0.45,angle=90]{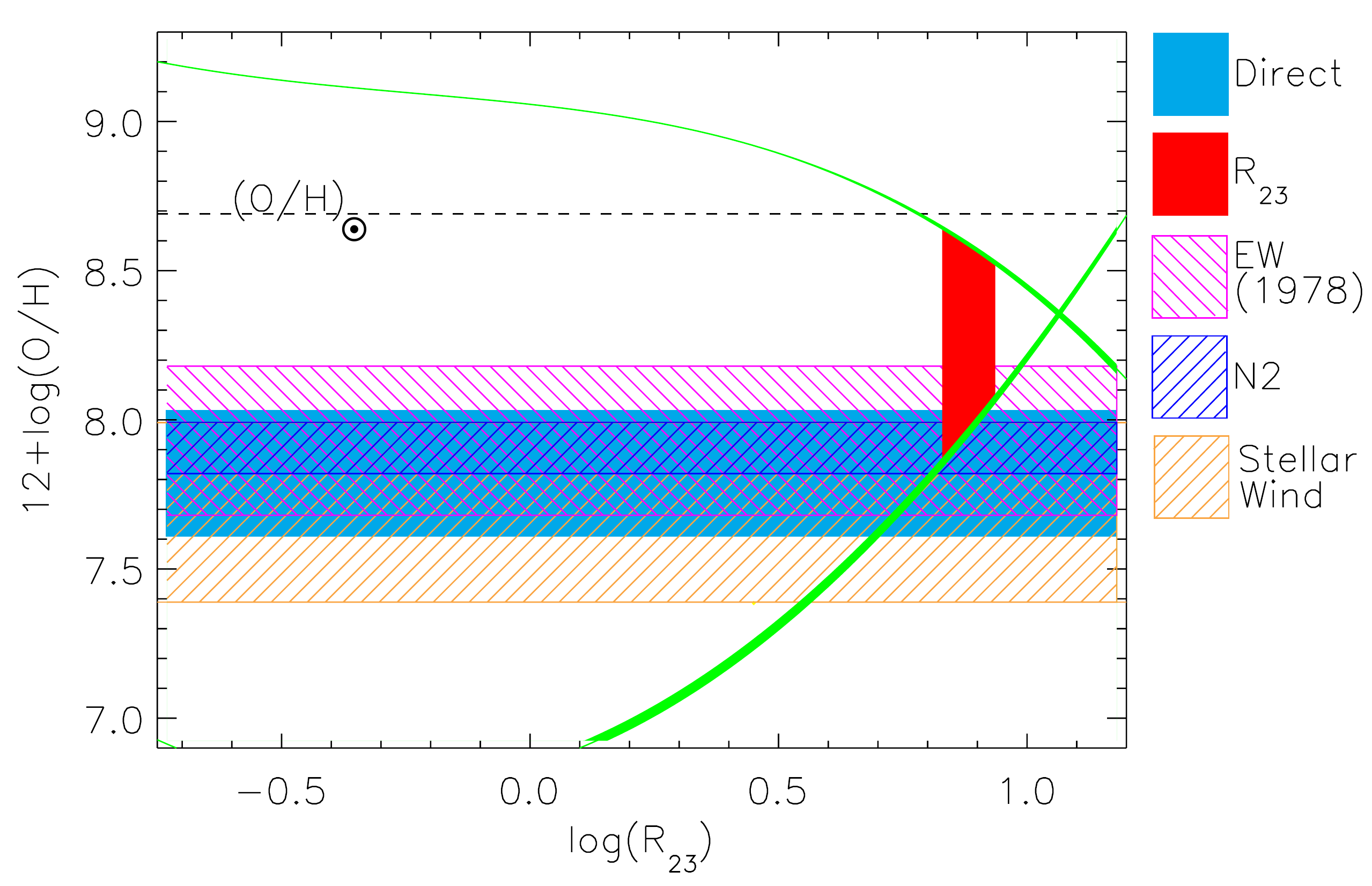}
\caption{Comparison of different abundance indicators in CSWA\,20.
The green continuous lines show the upper and lower branch
relations between the $R23$ index and the oxygen abundance 
(Section~\ref{sec:R23}); the red vertical band connects the two
solutions admitted by the $R23$ index measured in CSWA\,20.
The other methods are indicated by the colour key on the right.
Note that although the $y$-axis indicates the abundance
of oxygen, two of the methods [Stellar Wind and EW(1978)]
measure a mix of elements [mostly Fe in the case of EW(19878)]. 
In plotting them here, we have assumed solar  
abundances relative to oxygen for the elements probed.
The values shown graphically in the Figure are
collected in Table~7.}
 \label{fig:metals}
\end{figure*}

\begin{table*}
 \begin{minipage}{170mm}
 \caption{Summary of metallicity determinations$^{a}$.}
 \begin{tabular}{lllll}
\hline
\hline
Method & Element(s) & $Z/Z_{\odot}^{\rm b}$  & Range of $Z/Z_{\odot}^{\rm c}$& Comments \\ 
\hline
$T_{\rm e}$   & O                   & 0.14                    & 0.08--0.22  & \hii\ regions \\
$R23$              & O                   & 0.19                    & 0.14--0.24  & \hii\ regions \\
$N2$               & O                   & 0.17                     & 0.13--0.20  & \hii\ regions \\
C\,{\sc iv}     & C, N, O, Fe   & \ldots                & 0.05--0.20  & Stellar wind, O stars\\
1978                & Fe                   & 0.17                    & 0.10--0.31  & Photospheric, B stars \\
\hline
 \end{tabular}

$^{\rm a}$ Solar abundances from \citet{Asplund:2009}.\\
$^{\rm b}$ Best fitting value of metallicity relative to solar
(on a linear scale).\\
$^{\rm c}$ Range of metallicity relative to solar
(on a linear scale).
\end{minipage}
\label{tab:metals}
\end{table*}

\subsection{The 1978  index}
\label{sec:1978index}

Between $\sim 1900$ and 2050\,\AA\ in the 
UV spectrum of a stellar population, there is a 
broad blend of \feiii\  lines formed
in the photospheres of B-type stars. 
\citet{Rix:2004} gave an account of the historical background
to using this blend as an indicator of the Fe abundance.
The main drawback in its application to high-$z$ galaxies 
is the shallow nature of this spectral feature,
whose measured strength is therefore very 
sensitive to the exact placement of the continuum level.

We used the latest version of the Starburst99+WM-Basic
models \citep{Leitherer:2010} to recalibrate the `1978 index'
(defined as the equivalent width between 1935  and 2020\,\AA)
of \citet{Rix:2004} who used an earlier version of the WM-Basic
code \citep[see][for a description of the update]{Leitherer:2010}. 
We followed the steps prescribed by
\citet{Rix:2004}: we convolved the output spectra of Starburst99
(again for the case of 100\,Myr old, continuous star formation with a Salpeter IMF)
with a Gaussian of ${\rm FWHM} = 2.5$\,\AA\ to match
the resolution of most spectra of high-$z$ galaxies in the literature,
and normalised the spectra by dividing 
by a spline fit to a set of pseudo-continuum points \citep[Table~3 of][]{Rix:2004}.
The left panel of Figure~\ref{fig:1978index} compares the new
metallicity calibration of the 1978 index with
the earlier one by \citet{Rix:2004}; 
the behaviour of the index
is similar, but the new values are lower by $\sim 0.5$--1\,\AA\
for a given metallicity.  The new dependence of 
EW(1978) on metallicity (essentially [Fe/H])
can be approximated by the quadratic
relation:
\begin{equation}
\log\left(\frac{Z}{\Zsol}\right) = 0.03 x^2 + 0.17 x  -1.59
\label{eq:1978}
\end{equation}
where $x =$\,EW(1978) within the range 1.4--5.4\,\AA. 

As can be seen from Figure~\ref{fig:1978index} (right panel), the
combination of the WM-basic libraries of theoretical 
stellar spectra and the 
Starburst99 spectral synthesis code  is successful
at reproducing the complex blend of 
photospheric absorption lines in this portion
of the UV spectrum. Again, the observed spectrum
of CSWA\,20 appears to be intermediate between
the 0.05 and 0.2\,\Zsol\ cases. 
We measure EW(1978)$_{\rm CSWA\,20} = 3.1 \pm 0.7$\,\AA;
according to equation~(\ref{eq:1978})
this value corresponds to a metallicity
$\log(Z_{\rm CSWA\,20}/Z_\odot) = -0.76 \pm 0.25$
($Z_{\rm CSWA\,20} \simeq 1/3$--$1/10$\,\Zsol).

\section{Discussion}
\label{sec:discuss}


\begin{figure}
\includegraphics[scale=0.4]{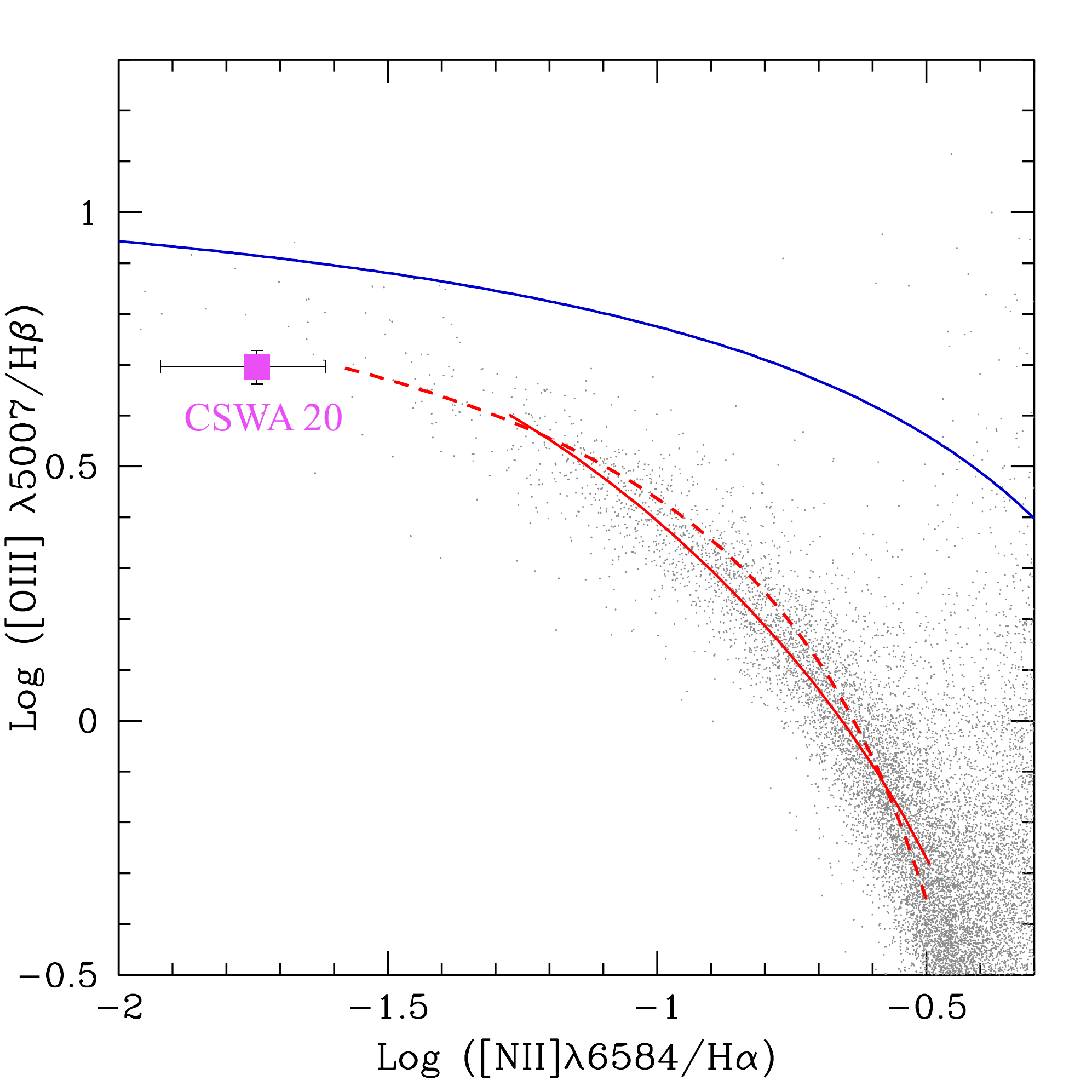}
\caption{Zoom-in on a portion of the 
BPT diagram of \citet{Baldwin:1981}, showing
the location of CSWA\,20 (magenta point).
The small grey points are galaxies from the 
Sloan Digital Sky Survey; the red curves are
fits to the locus of local galaxies based on the
relations by \citet{Maiolino:2008} and \citet{Kewley:2008}.
The blue continuous line divides star-forming galaxies
from Active Galactic Nuclei according to the models
by \citet{Kewley:2001}.
}
 \label{fig:BPT}
\end{figure}

In Table~7 and Figure~\ref{fig:metals} we bring together 
the different determinations of metallicity in CSWA\,20
discussed in Sections~\ref{sec:abund} and \ref{sec:indicators}.
Evidently, in this galaxy all five methods give consistent values
of the metallicity. The agreement between
the `direct' method, based on detecting weak auroral lines, 
and the more approximate strong-line indices $R23$ and
$N2$ is encouraging.
On the other hand, and perhaps counterintuitively, 
the apparent good match between the abundance
of O in the \hii\ regions of  CSWA\,20 
and that of Fe in the photospheres
of its early-type stars may be less reassuring. 
In Section~\ref{sec:C/O} it was suggested that the unusually
low C/O ratio may be related to a rapid 
timescale of chemical enrichment, with the release of
C into the ISM somewhat delayed relative to 
that of O. If this explanation is correct, then one
may also have expected a subsolar Fe/O ratio, by
a factor of $\sim 2$--3. 
Quite possibly, such a degree
of $\alpha$-element enhancement
is hidden in the uncertainties 
associated with the calibration and use
of stellar metallicity indicators.
It may well be premature to try and discern
such subtleties  in the chemical composition
of high-$z$ galaxies, given the limitations of
current models and data.

Returning to the nebular emission lines,
it is now well established that high-$z$ 
star-forming galaxies are clearly separated 
from local counterparts when their emission
line ratios are plotted on the plane
\foiii/\hb--\fnii/\ha, commonly referred to
as the `BPT' diagram of \citet{Baldwin:1981}.
The recent work by Steidel et al. (in preparation)
provides a vivid demonstration of the
shift to higher values  of  \foiii/\hb\
for a given \fnii/\ha\ (and vice versa)
found in  galaxies at $z \sim 2.3$, compared
to the bulk of star-forming galaxies
in the nearby Universe
\citep[see also][and references therein]{Brinchmann:2008b, Liu:2008, Masters:2014}.
Such a shift has been interpreted as the result of 
more extreme conditions (higher densities, more intense and 
harder radiation fields) in the \hii\ regions of
the luminous high-$z$ galaxies targeted by 
existing surveys \citep[e.g.][]{Kewley:2013}.

It is reasonable to question, therefore, 
whether calibrations of the strong-line indices 
developed from samples of nearby galaxies
can be applied in the high-redshift Universe,
and what systematic errors they may introduce
in the abundances so derived. 
As can be seen from Figure~\ref{fig:BPT},
CSWA\,20 lies near the top-left corner of the BPT diagram,
at log(\foiii/\hb), log(\fnii/\ha)\,$ = +0.69, -1.75$,
a locus occupied by few galaxies at all redshifts
(its weak \fnii\ lines are below the detection
limits of most large surveys).
Nevertheless, it does appear to lie on an extrapolation
to low metallicity of the mean relation for 
local galaxies. Thus, it may not be surprising 
that all three measures of the oxygen abundance in 
Table~7 are in good mutual agreement,
and we gain little guidance from CSWA\,20 
in assessing the systematic errors which may
affect the application of the strong-line indices to 
other star-forming galaxies at $z = 2$--3.


\begin{figure}
\vspace*{0.2cm}
\includegraphics[scale=0.4]{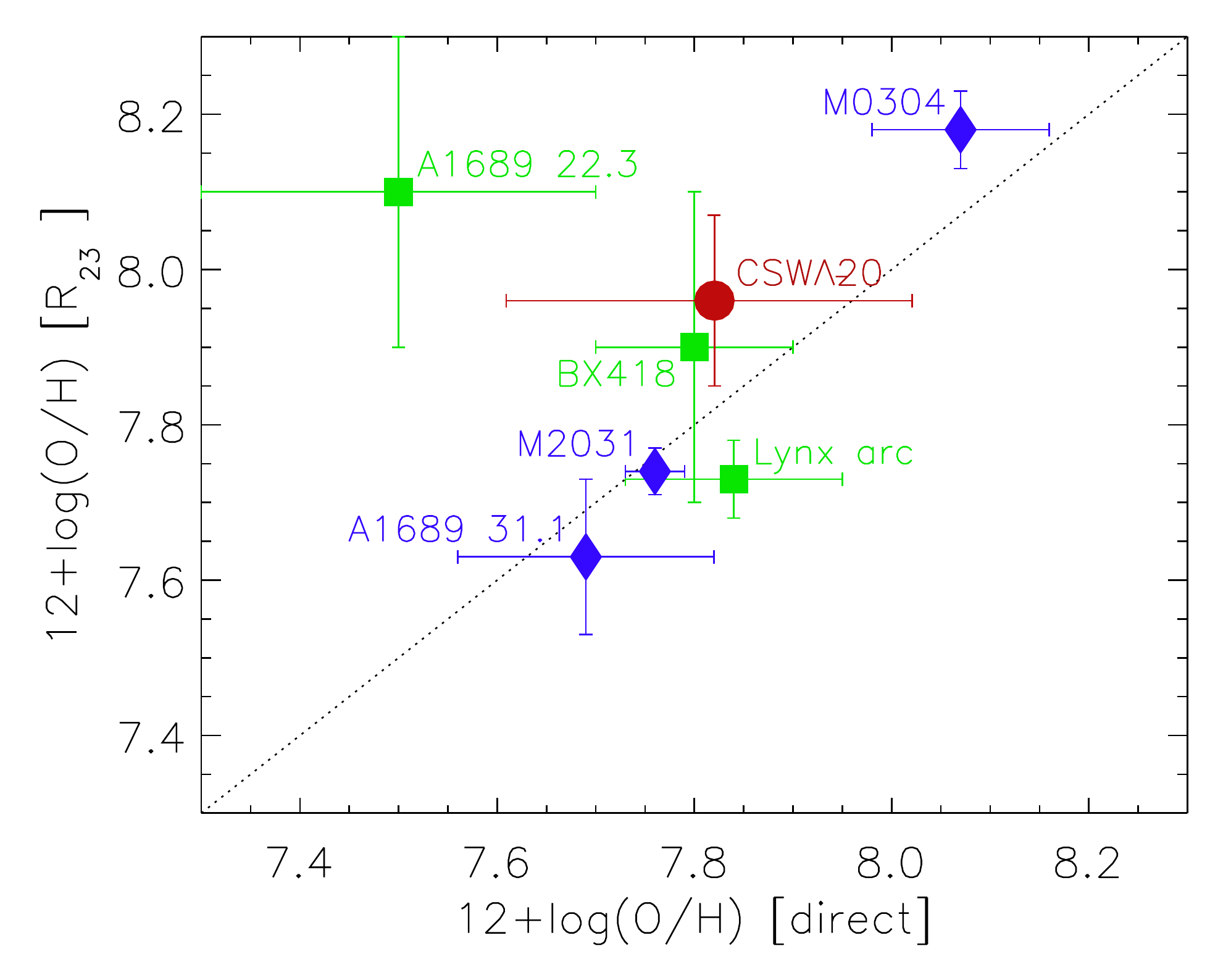}
\caption{Comparison between the oxygen abundance deduced 
from auroral lines (the `direct' method) and from the 
strong-line index $R23$ in the few high-$z$ star-forming
galaxies where both measures are available from the literature.
The blue diamonds are from the work by \citet{Christensen:2012b}.
The green squares are from \citet[][A1689 22.3]{Yuan:2009},
\citet[][BX418]{Erb:2010}, and \citet[][Lynx arc]{Villar:2004}, respectively.
$1 \sigma$ uncertainties are indicated by the error bars.
}
 \label{fig:R23vTe}
\end{figure}

To explore this point further,
we compare in Figure~\ref{fig:R23vTe} the oxygen abundance
deduced with the $T_{\rm e}$ and $R23$ methods 
in the few high-$z$ star-forming galaxies
where both values have been reported in the 
literature.\footnote{We note here that the reservations
expressed by \citet{Stasinska:2005} concerning biases
in the $T_{\rm e}$ method do not apply at
the low metallicities of these galaxies.}
Evidently, within the limitations of this very small sample, 
CSWA\,20 is not unique in showing a reasonable
agreement between the two approaches, with the only deviant point
being the $z = 1.705$ galaxy 22.3 behind the cluster Abell~1689
observed by \citet{Yuan:2009}\footnote{We note with some concern, however,
that these authors measured a large redshift difference
(corresponding to $\sim 1100$\,km~s$^{-1}$)
between the feature they identified as \foiii\,$\lambda 4363$ 
and all the other emission lines in this galaxy.}. 
Future near-IR observations of much larger samples
of $z \simeq 1.5$--3 star-forming galaxies, now made possible 
by the advent of the MOSFIRE spectrograph
\citep{McLean:2010}, will undoubtedly 
help us assess better the validity of different
nebular abundance measures in the high-$z$ galaxy population.

\section{Summary}
\label{sec:summary}

In this paper, we have analysed VLT+X-shooter observations
of CSWA\,20, a $z = 1.433$ star-forming galaxy
whose apparent magnitude  is boosted by a factor
of 11.5 ($-2.7$\,mag)  by  gravitational lensing 
by a foreground massive elliptical galaxy at $z =  0.741$.
The boost provided by the lens gives us access to weaker
spectral features and allows us to record the spectrum
at higher  resolution
than is normally the case;
coupled with the wide wavelength
coverage of X-shooter, all these factors 
give us an
in-depth view of the physical conditions in a 
galaxy which was actively forming stars $\sim 9$\,Gyr ago.
The main results of this work are as follows.

1. CSWA\,20  has a metallicity $Z \sim 1/7$\,\Zsol,
measured from its \hii\ regions and massive stars.
Its current star-formation rate is $\sim$6\,\Msol~yr$^{-1}$, one order of magnitude
lower than the star formation rate of
$L^\ast$ galaxies at $z \sim 2$ \citep[][]{Reddy:2008}.
The sub-solar metallicity is reflected in the high temperature
of its \hii\ regions, \elt\,$\simeq 17000$\,K, 
compared to \elt\,$\simeq 10000$\,K typical of \hii\
regions in nearby galaxies.

2. Carbon is significantly less abundant than oxygen,
compared to the solar ratio of of these two elements.
The value we find, $\log {\rm C/O} = -1.06 \pm 0.32$,
is among the lowest measured in metal-poor \hii\ galaxies,
and $\sim 2$--3 times lower than in Galactic halo
stars of similar oxygen abundance.
Possibly, chemical enrichment has proceeded at
such a rapid rate in CSWA\,20 that the low and intermediate
mass stars which contribute some of the carbon
have not had sufficient time yet to catch up with
the prompt release of oxygen from more massive stars.

3. With our data, we can use five different methods  
to measure the metallicity of the galaxy; three are
based on emission lines from \hii\ regions and
two on absorption features in the atmospheres of
massive stars. All five routes give the same value
of metallicity within a factor of $\sim 2$.
Of the three determinations of the 
oxygen abundance from emission lines,
one uses the
density and temperature of the \hii\
regions, while the other two rely
on approximate local calibrations of  
strong line ratios. In CSWA\,20, all
three methods give the same answer,
but this may not be surprising, given that
this galaxy falls on an extrapolation 
to low metallicity of the locus occupied
by local galaxies in the BPT diagnostic diagram.

4. In the course of the analysis performed here,
we have recalibrated the dependence on metallicity
of the \feiii\ 1978 index of \citet{Rix:2004},
using the latest combination of the Starburst99
spectral synthesis code and the WM-basic suite of 
theoretical stellar spectra.

5. Most interstellar absorption lines are unusually weak
in CSWA\,20, probably as a result of a very patchy 
coverage of the starburst by the ISM, as viewed from our direction.
The resolved kinematics of the absorption
reveal gas spread over nearly 1000\,km~$^{-1}$,
with most of the absorption taking place in
gas which is outflowing from the galaxy with
velocities of up to $\sim 750$\,km~s$^{-1}$.
The maximum speed reached by the outflow
is similar to the values measured in a few other
gravitationally lensed galaxies at $z \sim 1.5$--3,
despite the order-of-magnitude difference in the
star-formation rates among these galaxies.
Thus it is unclear to what extent the terminal
velocity of the `superwind'  in galaxies 
with high star-formation rate surface densities
depends on the total energy and momentum deposited
into the ISM by the central starburst.

All in all, our observations of CSWA\,20 have left us with
many questions, as well as providing some answers.
While it is generally appreciated that physical conditions
in the galaxies responsible for the bulk of the 
star formation activity at earlier epochs are markedly different from
those encountered in the local Universe, we still lack
a proper physical framework that can help us put into context
the results of studies such as the one presented here.
A major limiting factor at present is the paucity of 
high redshift galaxies bright enough to be studied in 
detail, so that correlations between age, mass, 
star-formation history, metallicity,
and large-scale outflows in the ISM can be explored.
Undoubtedly, future observations of other gravitationally
lensed galaxies will in time fill in the many gaps that
still remain.

\section{Acknowledgements}
We are grateful to the European Southern Observatory time assignment 
committee who awarded time to this programme, 
and to the staff astronomers at Paranal who conducted the observations.
It is a pleasure to acknowledge the help by
Sergey Koposov with various aspects of the data analysis.
Vasily Belokurov kindly provided the WHT images of 
CSWA\,20 reproduced in Figure~\ref{fig:CSWA20}, and
Naveen Reddy  the composite of 
BM galaxy spectra used in Figure~\ref{fig:BMcomp}.
We are also grateful to 
Mike Barlow, Claus Leitherer and Daniela Calzetti for useful discussions, and to the anonymous referee whose comments and suggestions improved the paper. .
LC is supported by the European Union under a
Marie Curie Intra-European fellowship, contract PIEF-GA-2010-274117.


\label{lastpage}

\end{document}